\documentclass[aps,pra]{revtex4-1}
\usepackage{amsmath}
\usepackage{amsfonts}
\usepackage{amssymb}
\usepackage{mathtools}
\usepackage{graphicx}
\usepackage{tabularx}
\usepackage[caption=false,position=b]{subfig}
\usepackage[colorlinks]{hyperref}
\usepackage[capitalize]{cleveref}
\usepackage{wrapfig}

\usepackage{yypreamble}

\usepackage{color}

\newcommand{\vm}{\vec{m}}
\newcommand{\vn}{\vec{n}}

\newcommand{\Ns}{\mathcal{N}}
\newcommand{\Ms}{\mathcal{M}}
\newcommand{\J}{\mathcal J}

\begin{document}

\title{Reservoir engineering with localized dissipation:  dynamics and pre-thermalization}
\author{Yariv Yanay}
\email{yariv@lps.umd.edu}
\affiliation{Laboratory for Physical Sciences, 8050 Greenmead Dr., College Park, MD 20740}
\author{Aashish A. Clerk}
\affiliation{Pritzker School for Molecular Engineering, University of Chicago, 5640 S. Ellis Ave., Chicago, IL 60637}

\date{\today}


\begin{abstract}
Reservoir engineering lattice states using only localized engineered dissipation is extremely attractive from a resource point of view, but can suffer from long relaxation times.  Here, we study the relaxation dynamics of bosonic lattice systems locally coupled to a single squeezed reservoir. Such systems can relax into a highly non-trivial pure states with long-range entangement \cite{Yanay2018}.  In the limit of large system size, analytic expressions for the dissipation spectrum can be found by making an analogy to scattering from a localized impurity.  This allows us to study the cross-over from perturbative relaxation to a slow, quantum-Zeno regime.  We also find the possibility of regimes of accerelated relaxation due to a surprising impedance matching phenomena.  We also study intermediate time behaviours, identifying a long-lived ``prethermalized" state associated that exists within a light cone like area.  This intermediate state can be quasi-stationary, and can very different entanglement properties from the ultimate dissipative steady state.
\end{abstract}

\maketitle


\section{Introduction}

In quantum information applications, states with entanglement and other non-classical properties serve as basic resources. A powerful approach to generating such states is to couple the system of interest to a tailored dissipative environment, so that the resulting system steady state is the desired non-trivial quantum state; this approach is known as  reservoir engineering \cite{Poyatos1996,Plenio2002}. There is by now a growing body on work on utilizing such techniques, ranging from the stabilization of systems with a few degrees of freedom (see e.g. Refs.~\onlinecite{Krauter2011,Murch2012,Lin2013,Shankar2013,Leghtas2015,Wollman2015}), to the preparation of many body states via controlled system-wide dissipation (see e.g. Refs.~\onlinecite{Diehl2008,Verstraete2008,Kraus2008,Cho2011,Koga2012,Ikeda2013,Quijandria2013,Ticozzi2014}).  Work has also showed that in some cases many-body states can be stabilized by controlling dissipation in a limited, localized spatial region only \cite{Zippilli2015,Ma2016,Ma2017,Ma2017a}.  Such an approach was recently implemented in a superconducting circuit experiment to stabilize a bosonic Mott insulator \cite{Ma2019}.    Our recent work in Ref.~\cite{Yanay2018} analyzed a particularly striking example of this local approach to reservoir engineering:  it  demonstrated that an entire class of free boson lattice systems with a generalized chiral symmetry can be stabilized in this manner by making use of the lattice symmetry and a single,  localized,  squeezed dissipative reservoir.  This allows the stabilization of non-classical, often highly non-locally entangled Gaussian pure states \cite{Yanay2018}. 

While the ability to stabilize non-trivial states using a localized coupling to engineered dissipation is extremely attractive in terms of the needed experimental resources, an obvious potential drawback is that the timescale for relaxing to the steady state can be extremely long (making one more susceptible to unwanted dissipation and decoherence).    The relevant relaxation time in such schemes typically scales with system size.  Intuitively, this makes sense:  correlations are generated locally at the dissipative site, and take time to propagate throughout the lattice. As a result, a long intermediate time regime exists before the final steady state is reached.  Understanding this dynamical phenomena is of interest for many reasons.  It would allow one to optimize the relaxation process and speed when local dissipation is used in reservoir engineering. In addition, the system's path from its initial configuration to the dissipative steady state may exhibit different qualities and new physics that is present in neither.

In this article we study the relaxation dynamics in this kind of reservoir engineering setup.  We focus on the simple but paradigmatic case considered in Ref.~\cite{Yanay2018}:  a bosonic lattice system described by a quadratic hopping Hamiltonian, coupled to a Markovian reservoir on just a single site.  As we show, the relaxation here can be fully characterized by the eigenvalues of the system's dynamical matrix.  Further, the corresponding relaxation rates have a strong parallel to the physics of scattering off a localized potential and the Friedel sum rule.   Analyzing the behavior of these modes as local dissipation strength is increased, we see a transition from perturbative relaxation of the system's original eigenstates to a quantum Zeno-like decoupling of the dissipative site.  While these limiting cases could easily be anticipated, we also find that under some circumstances, there can be a surprising resonant enhancement of mode relaxation rates for intermediate coupling strengths; this can be interpreted as an impedance matching phenomenon.  We also consider the overall evolution of the system from its initial state, describing the intermediate-time behavior in large systems.  We find the existence of a quasi-steady state, whose form and duration is dictated by ballistic propagation physics.  This quasi-steady state describes sites where correlations have had time to propagate from the dissipative site, but have not had time to reach the system boundary and then return.  Surprisingly, we find that this quasi-steady state can have a very different form and pattern of entanglement than the final steady state of the system.  

The remainder of the article is organized as follows. In \cref{sec:model} we outline the basic model of the class of dissipative systems we consider. In \cref{sec:modes}, we find the eigenmodes of our system's dynamical matrix and corresponding ``dissipation spectrum", and analyze its behavior. We provide several numerical examples and one analytical solution of a sample system. Finally, in \cref{sec:intertime}, we turn to the intermediate-time behavior of the system, where the state differs both from the initial state and the final stabilized state.  We show the emergence of a quasi-steady state within the light cone corresponding to ballistic propagation of correlations from the dissipative site.

\section{Model\label{sec:model}}

We consider the same class of model studied in our previous work \cite{Yanay2018}, though to start, do not impose any sort of symmetries. We consider a generic particle conserving quadratic bosonic Hamiltonian,
\begin{equation}\begin{split}
\hat \H & =  \vec{\hat a}\dg\cdot H \cdot \vec{\hat a} = \sum_{\mathclap{\vm, \vn}}H_{\vm,\vn}\hat a_{\vm}\dg\hat a_{\vn} 
\label{eq:HS}
\end{split}\end{equation}
where $\hat a_{\vn}$ ($\hat a_{\vn}\dg$) is the annihilation (creation) operator for a boson on site $\vn$, and the Hamiltonian matrix $H$ consists on-site potentials $H_{\vn,\vn} = V_{\vn}$ and hopping elements $H_{\vm,\vn} = J_{\vm,\vn}$. The site labels $\vn$ describe an arbitrary $d$-dimensional lattice with $N$ sites, and we do not assume any symmetry or translational invariance to begin with. This Hamiltonian describes a range of bosonic systems, including coupled arrays of superconducting cavities or mechanical oscillators.

We linearly couple a single ``drain'' site, marked by $\vn_{0}$ to a Markovian, Gaussian reservoir at zero temperatrure. Using standard input-output theory \cite{Gardiner2004}, the Heisenberg-Langevin operator equations of motion are then
\begin{equation}
	\dot{\hat a}_{\vn}(t) = -i\br{\hat a_{\vn}(t),\hat \H} - \gd_{\vn,\vn_{0}}\mat{\half[\Gamma]\hat a_{\vn_{0}}(t) - \sqrt{\Gamma}\hat \zeta(t)},
\label{eq:evolreal}
\end{equation}
where $\Gamma$ is a dissipation rate parameterizing the strength of the coupling to the reservoir and $\hat \zeta(t)$ is a Gaussian white noise operator describing its  vacuum fluctuations. We will eventually take $\hat\zeta(t)$ to describe squeezed vacuum fluctuations, but for now keep things general (as the system's characteristic relaxation rates are independent of the nature of the noise).

The Hamiltonian of \cref{eq:HS} can also be written in diagonal form using its energy eigenmodes $\hat{b}_{i}$ and corresponding eigenstate wavefunctions $\psi_{i}\br{\vn}$:
\begin{equation}
\hat \H = \sum_i \gve_{i}\hat b_{i}\dg\hat b_{i}, \qquad \hat b_{i}\dg = \sum_{\vn}\psi_{i}\br{\vn}\hat a_{\vn}\dg.
\end{equation}
Without loss of generality, we label these energy eigenmodes so that $\gve_{i+1}\ge \gve_{i}$. We will also take the spectrum to be non-degenerate with $\psi_{i}\br{\vn_{0}} \ne 0$ for all modes; in other words, we focus on the portion of the spectrum that is coupled to the drain. Any other modes are unaffected by the dissipation dynamics (see discussion in \cite{Yanay2018}).

In the eigenmode basis, including the coupling to the reservoir, the operator equations of motion take the form
\begin{equation}
	\dot{\hat b}_{i}(t) = -i \sum_{j} A_{ij}\hat b_{j}(t) + e^{-i\varphi_{i}}\sqrt{\bar \Gamma_{i}}\hat\zeta(t), 
\label{eq:evolbe}
 \end{equation}
where the dynamical matrix $A$ is given by
\begin{equation}
	A_{ij} = \gd_{i,j}\gve_{i} - i e^{i\p{\varphi_{j}-\varphi_{i}}}\half \sqrt{\bar \Gamma_{i}\bar \Gamma_{j}}.
\label{eq:Amat}
\end{equation}
Here
\begin{equation}
\varphi_{i} = \arg \br{ \psi_i\br{\vn_{0}} }, \qquad \bar \Gamma_{i} = \abs{\psi_{i}\br{\vn_{0}}}^{2}\Gamma
\label{eq:wftoGi}
\end{equation}
are the phase and magnitude of the coupling between the energy eigenmode $i$ and the reservoir.

\section{Dynamical modes and characteristic dissipation rates \label{sec:modes}}

For a generic Markovian system, it is common to characterize relaxation time scales by considering the the eigenvalues of the Liouvillian which governs the evolution of the system's reduced density matrix \cite{Prosen2010}.  Here, the linearity of our system makes life much easier.  We can fully characterize the system's dynamics and relaxations by simply diagonalizing the dynamical matrix $A$ defined in \cref{eq:Amat}. Its eigenvalues are the characteristic mode frequencies of the linear dynamics of \cref{eq:evolbe}. The non-Hermitian nature of $A$ means these mode frequencies will be complex, with non-positive imaginary part corresponding to loss. We therefore calculate the left-eigenvectors $V_{ij}$ and eigenvalues $\gl_{i}$ of $A$, 
\begin{equation}
\sum_{j}V_{ij} A_{jl} = \gl_{i}V_{i,l}.
\label{eq:lefteigenAdef}
\end{equation}
As shown in \cref{app:eigens}, the left-eigenvectors are given by
\begin{equation}
V_{ij} = \frac{1}{2}\frac{e^{i\varphi_{j}}\sqrt{\bar\Gamma_{j}}}{\gl_{i} - \gve_{j}},
\label{eq:eigenA}
\end{equation}
and its eigenvalues $\gl_{i}$ are the solutions of the self-consistency equation
\begin{equation}
S\p{\gl} = \frac{1}{2} \sum_{j}\frac{ \bar\Gamma_{j} }{\gl - \gve_{j}} = i.
\label{eq:consist}
\end{equation}

These eigenvectors define a set of operators whose time evolution can be immediately integrated, 
\begin{equation}
\tilde b_{i} = \sum_{j}V_{ij}\hat b_{j}, \qquad \dot{\tilde b}_{i}(t) = -i\gl_{i}\tilde b_{i}(t) + i\hat\zeta(t).
\label{eq:evoleigen}
\end{equation}
Note that these operators are not a set of independent, canonical annihilation operators, but instead satisfy
\begin{equation}
\br{\tilde b_{i},\tilde b_{j}\dg} = \sum_{l}V_{il} V_{jl}^{*} = \frac{i}{\gl_{j}^{*}-\gl_{i}}.
\end{equation}
Despite this, they will be useful in the analysis which follows.

\subsection{Dissipation Spectrum\label{sec:spectrum}}

Our goal is to use \cref{eq:consist} to calculate the dynamical matrix eigenvalues $\lambda_i$ in the limit of a large system.  Note that the evolution described in \cref{eq:evolreal} is analogous to scattering off a localized impurity, except now, the impurity potential is imaginary (i.e.~effectively non-Hermitian).  Thus, similar to the standard treatment of potential scattering \cite{Landau1977},  we look for dynamical eigenvalues $\gl_i$ that are just a small shift of the original energy eigenvalues $\gve_i$,
\begin{equation}
	\gl_{i} = \gve_{i} + \gd\gl_{i}.
\end{equation}
The real and imaginary parts of this shift corresponds to an energy shift $\delta \nu_{i}$, and an inverse lifetime $\gamma_{i}/2$,
\begin{equation}
	\gd\gl_{i} = \gd\nu_{i} - i \half[\gamma_{i}].
\end{equation}
The sum in \cref{eq:consist} can then be rewritten
\begin{equation}
	S\p{\gve_{i} + \gd\gl_i} = \frac{1}{2} \sum_{n}\frac{ \bar\Gamma_{i+n}}{\gd\gl_i - \p{\gve_{i+n} - \gve_{i}}}.
\label{eq:consistshift}
\end{equation}

In the case of a large system, we generally expect the shift and the relaxation rate to be quite small, ${\abs{\gd\gl_{i}} \sim 1/N}$. The sum in \cref{eq:consistshift} is then dominated by the resonant terms, $\abs{\gve_{i+n} - \gve_{i}}\sim \abs{\gd\gl_i}$. Further, we generically expect the unperturbed density of states and the coupling to the drain, $\bar \Gamma_{i+n}$, to remain roughly constant for these resonant contributions. We thus approximate the spectrum and the local coupling for modes near $i$ as
\begin{equation}
\bar \Gamma_{i+n} \to \bar \Gamma_{i}, \qquad \gve_{i+n} \to \gve_{i} + n \Delta_{i},
\label{eq:nearppx}
\end{equation}
where $\Delta_{i} = \p{\gve_{i+1} - \gve_{i-1}}/2$ is the energy spacing near $\gve_{i}$. We can then approximate \cref{eq:consistshift} as this resonant portion,
\begin{equation}\begin{gathered}
S\p{\gve_{i} + \gd\gl_i} \approx \tilde S_{i}\p{\gd\gl_i}
	\equiv \frac{1}{2} \sum_{\mathclap{n=-\infty}}^{\infty} \frac{ \bar \Gamma_{i}}{\gd\gl_i - \Delta_{i}n}.
\label{eq:Si}
\end{gathered}\end{equation}
This is equivalent to assuming a constant density of states, and eigenmode wavefunctions that have a constant amplitude at the drain site.  

The sum in \cref{eq:Si} can be immediately evaluated to find
\begin{equation}\begin{gathered}
\gd\gl_{i} \approx -i\tfrac{\Delta_{i}}{2\pi}\ln\br{\tfrac{\tfrac{\Delta_{i}}{\pi} + \half[\bar \Gamma_{i}] }{\tfrac{\Delta_{i}}{\pi}- \half[\bar \Gamma_{i}]}}.
\label{eq:gammaappx}
\end{gathered}\end{equation}
This is the usual expression for an s-wave scattering phase shift from a localized impurity potential \footnote{Recall that $\tan^{-1}x = \half[i]\log\tfrac{1-ix}{1+ix}$ and that the density of states $\rho_{i}\sim1/\Delta_{i}$}\cite{Mahan2000a}, with the potential taken to be imaginary, $V\to i\Gamma/2$. The phase shift determines the energy shift of the modes, and so the resulting imaginary portion gives the relaxation rate,
\begin{equation}\begin{gathered}
\gamma_{i} = -2\Im\br{\gd\gl_{i}} \approx \tfrac{\Delta_{i}}{\pi}
	\ln\abs{\tfrac{\tfrac{\Delta_{i}}{\pi} + \half[\bar \Gamma_{i}] }{\tfrac{\Delta_{i}}{\pi}- \half[\bar \Gamma_{i}]}}.
\label{eq:gammaappx}
\end{gathered}\end{equation}

Eq.~\ref{eq:gammaappx} is a key result of this paper.  It demonstrates that the relaxation rate associated with mode $i$ is controlled both by the local level spacing $\Delta_i$ of the unperturbed spectrum near $E = \gve_i$, as well as the coupling rate $\Gamma_i$ between mode $i$ and the localized dissipative reservoir.  $\Delta_i$ will generically scale as $1/N$; while the same is true for $\bar\Gamma_i$ if the mode $i$ is extended.  In this case the relaxation rate $\gamma_i$ will also scale as $1/N$.  For a localized eigenmode $i$, we instead expect  $\bar\Gamma_{i} \sim \exp\br{-r_{i}/\xi}$ for the mode's distance from the drain $r_{i}$ and its extent $\xi$.

We observe that the relaxation rates of the system are determined by the parameter $\bar\Gamma_{i}/\Delta_{i}$. This can be understood as 
the ratio of the mode's dwell time ($\Delta_{i}^{-1}$) to its dissipative lifetime ($\bar\Gamma_{i}^{-1}$).
We see two limits for this rate,
\begin{equation}
\gamma_{i} \propto \fopt{\bar\Gamma_{i} & \bar \Gamma_{i}\ll \Delta_{i}
\\  \Delta_{i}^{2}/\bar\Gamma_{i}& \bar \Gamma_{i}\gg \Delta_{i}.
}
\label{eq:gammaregimes}
\end{equation}
In the weak coupling limit, $\bar\Gamma_{i}\ll\Delta_{i}$, the original modes are perturbatively coupled to the dissipative bath, yielding  $\gamma_{i} \simeq \bar \Gamma_{i}$, a result that would be expected from a standard Fermi's Golden Rule calculation. In this limit the relaxation time of any localized mode grows exponentially with its distance from the coupled site. In contrast, for extended modes the dissipation rate will scale as $1/N$.  
The strong dissipation limit, $\bar\Gamma_{i}\gg\Delta_{i}$, can be understood in terms of quantum Zeno physics \cite{Misra1977}: the coupled site is measured by its bath faster than it can interact with these modes, and their dissipation is suppressed. 

It is also interesting to consider the behaviour of $\gamma_i$ in the regime where there is an approximate matching of timescales, $\bar \Gamma_i \sim \Delta_i$.  In this case, there is a form of impedance matching, as the propagation rate of waves arriving at the coupled site matches that of waves radiated into the bath. To understand the relaxation rates at this regime, we must take into account the non-resonant portions of the sum in \cref{eq:Si}. We define the remainder
\begin{equation}\begin{split}
R_{i}\p{\gd\gl} & 
	\equiv S\p{\gve_{i}+\gd\gl} - \tilde S_{i}\p{\gd\gl}
	=  \frac{1}{2}\sum_{\mathclap{\text{eigenvalues }j}}\frac{\bar\Gamma_{j}}{\gd\gl - \p{\gve_{j} - \gve_{i}}} - 
		\frac{1}{2}\sum_{\mathclap{n=-\infty}}^{\infty}\frac{\bar \Gamma_{i}}{\gd\gl - \Delta_{i}n} 
	\\ & = \frac{\gd\gl}{2}\br{\sum_{\text{e.v.\,}j}\frac{\bar\Gamma_{j}}{\gd\gl^{2} - \p{\gve_{j} - \gve_{i}}^{2}} - 
		\sum_{j=-\infty}^{\infty}\frac{\bar \Gamma_{i}}{\gd\gl^{2} -  \Delta_{i}^{2}\p{j-i}^{2}} } 
		+ \frac{1}{2} \sum_{j\ne i}\frac{\bar\Gamma_{j}\p{\gve_{j} - \gve_{i}}}{\gd\gl^{2} - \p{\gve_{j} - \gve_{i}}^{2}}.
\label{eq:Rdef}
\end{split}\end{equation}
This represents the correction to approximating the system as having constant wavefunctions and level spacing.

Recall that we expect $\gd \gl$ is extremely small in the limit of large system size:  $\gd\gl \sim \Delta_{i}\sim 1/N$.
If the approximations of \cref{eq:nearppx} hold, the resonant portions of the sums in the square brackets will cancel out, implying that the quantity inside the brackets is non-singular as $\gd \gl \rightarrow 0$.  The first term above then will be proportional to $\gd\gl$.  It follows that in this limit, $R_i$ will be dominated by the last term on the second line of Eq.~(\ref{eq:Rdef}).  Hence, to leading order in $1/N$:
\begin{equation}
	R_{i}\p{\gd\gl_{i}} \approx R_{i}\p{0} = \frac{1}{2}\sum_{j\ne i}\frac{\bar \Gamma_{j}}{\gve_{i} - \gve_{j}}.
	\label{eq:Riiappx}
\end{equation}
It follows that the only dependence on the strength of the dissipation $\Gamma$ in \cref{eq:Riiappx} is through an overall factor.  We thus define the rescaled remainder,
\begin{equation}
r_{i} \equiv \frac{\Delta_{i}/\pi}{\Gamma/2}R_{i}\p{0} = \frac{1}{\pi}\sum_{j\ne i}\frac{\Delta_{i}}{\gve_{i} - \gve_{j}}\abs{\psi_{j}\br{\vn_{0}}}^{2}.
\label{eq:ri}
\end{equation}
The dimensionless parameter $r_{i}$ is independent of the strength of the dissipation, and can be calculated for a given system and drain site. It characterizes how strong the corrections to the approximation in \cref{eq:nearppx} are for any particular mode $i$.

Note that while the resonant term $\tilde S_{i}$, as defined in \cref{eq:Si}, is determined purely by the effective dissipation and local level spacing for each energy eigenmode, the remainder $r_i$ depends on the global form of the entire spectrum, as well as fluctuations from mode to mode of the wavefunction at the drain-site. This means that the particular values of $r_{i}$ depend on the the specifics of the Hamiltonian involved. However, as $\sum_{j}\abs{\psi_{i}\br{\vn_{0}}}^{2} = 1$, 
we can interpret Eq.~(\ref{eq:ri}) as a weighted average of the terms in the summation. 
This allows it to capture the corrections from the non-uniformity of level spacing and drain-site wavefunctions over the entire spectrum.
Recalling that $\Delta_{i}$ is the energy spacing near $i$, we have $1 \le \abs{\gve_{j} - \gve_{i}}/\Delta_{i}$, we then have by definition $\abs{r_{i}}\le 1/\pi$. Furthermore, as the denominator of most terms in the sum is of the order of magnitude of the spectrum, we might expect $\abs{r_{i}}= O\p{1/N^{\ga}}$ for some positive $\ga$.

Combining \cref{eq:consistshift,eq:Si,eq:Rdef,eq:Riiappx,eq:ri}, we find the full expression for the relaxation rates of the eigenmodes,
\begin{equation}\begin{gathered}
\gamma_{i} = -2\Im\br{\gd\gl_{i}} \approx  \tfrac{\Delta_{i}}{\pi}
	\ln\abs{\tfrac{\tfrac{\Delta_{i}}{\pi} + \half[\Gamma]\p{\abs{\psi_{i}\br{\vn_{0}}}^{2} + i r_{i}}}{\tfrac{\Delta_{i}}{\pi} - \half[\Gamma]\p{\abs{\psi_{i}\br{\vn_{0}}}^{2} - i r_{i}}}}.
\label{eq:eigenvalues}
\end{gathered}\end{equation}
Note that if one knows the original system's eigenstate energies and wavefunctions at site $\vn_0$, then \cref{eq:eigenvalues} allows a calculation of the dissipation spectrum for an {\it arbitrary} dissipation strength $\Gamma$.  It thus allows calculation of the dissipation spectrum in a simple manner, without having to rediagonalize the dynamical matrix for each different choice of $\Gamma$.  
Further, for some systems it is possible to analytically calculate the spectrum and $r_{i}$; we provide such an example in the next section.  

\begin{figure}[ht] 
   \centering
   \parbox[c][0.22\textwidth][b]{0.05\textwidth}{
	\rotatebox{90}{$\gamma_{i}/\Delta_{i}$}
	}
   \subfloat[\label{fig:DisSpec-1D-side}1D chain, drain at end]
   {\parbox{0.25\textwidth}{
   	\parbox[c][0.2\textwidth][c]{0.2\textwidth}{\includegraphics[width=0.2\textwidth]{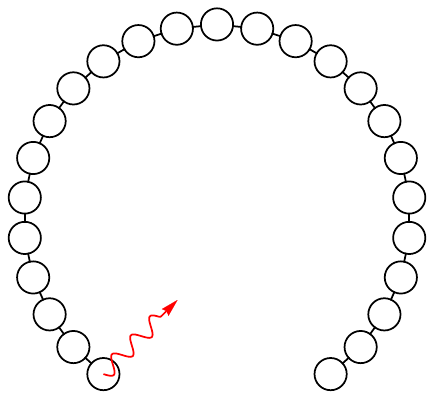}}
	
	\parbox[c][0.18\textwidth][c]{0.25\textwidth}{\includegraphics[width=0.25\textwidth]{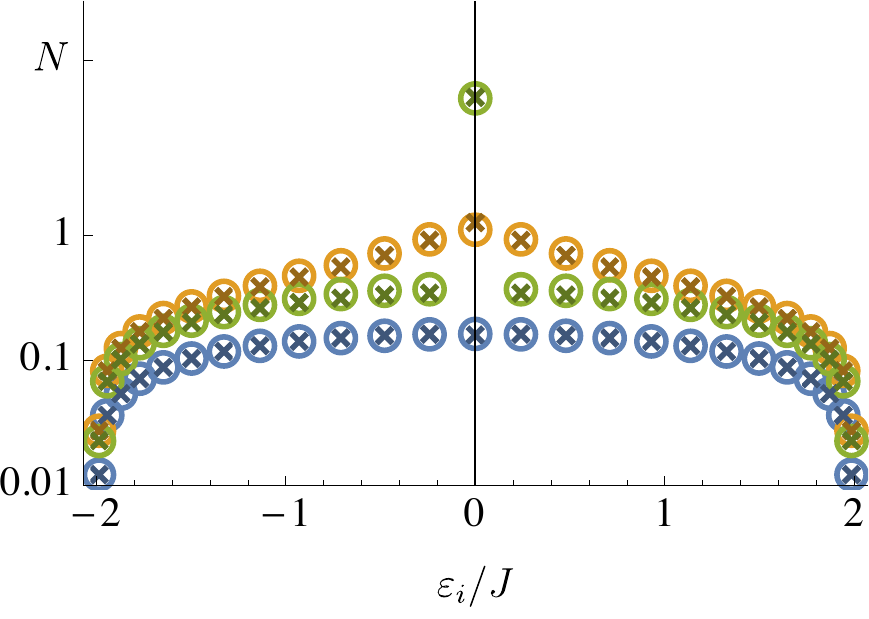}}
   }}
   \hfill
   \subfloat[\label{fig:DisSpec-1D-step}1D chain with step potential, drain at center]
   {\parbox{0.25\textwidth}{
   	\parbox[c][0.2\textwidth][c]{0.2\textwidth}{\includegraphics[width=0.2\textwidth]{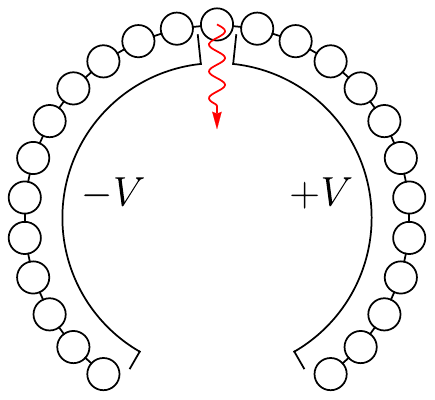}}
	
	\parbox[c][0.18\textwidth][c]{0.25\textwidth}{\includegraphics[width=0.25\textwidth]{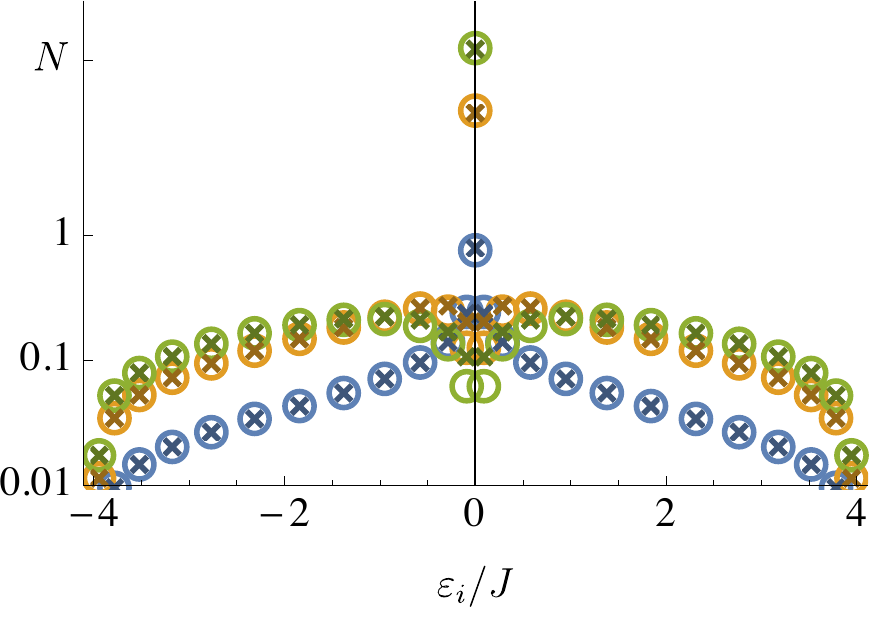}}
   }}
   \hfill
   \subfloat[\label{fig:DisSpec-Hoff}2D Hofstadter model with $\pi/2$ flux through each plaquette]
   {\parbox{0.25\textwidth}{
   	\parbox[c][0.2\textwidth][c]{0.2\textwidth}{\includegraphics[width=0.2\textwidth]{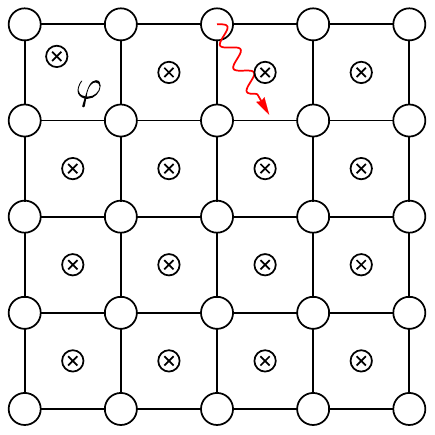}}
	
	\parbox[c][0.18\textwidth][c]{0.25\textwidth}{\includegraphics[width=0.25\textwidth]{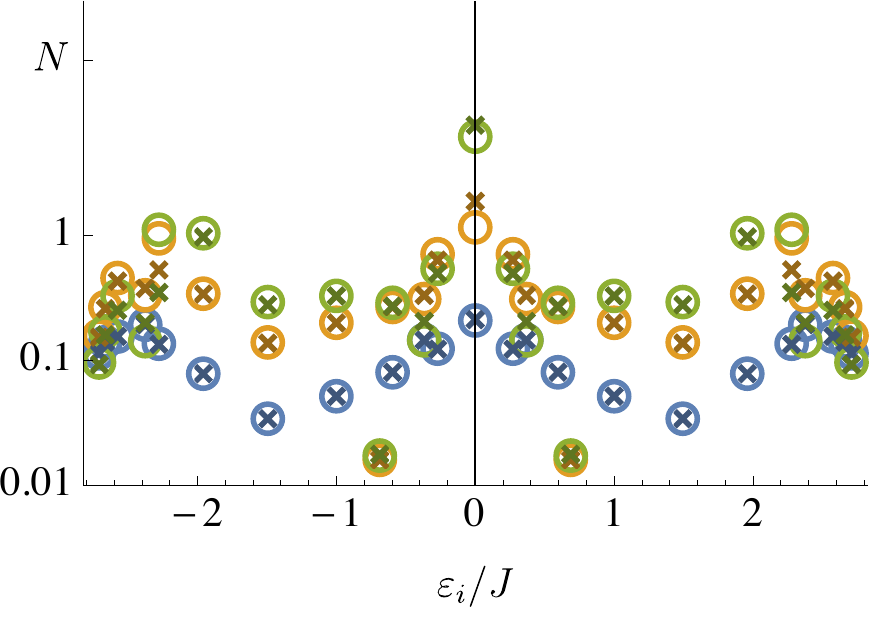}}
   }}
   \hfill
   \parbox[c][0.38\textwidth][b]{0.1\textwidth}{
	\includegraphics[scale=1]{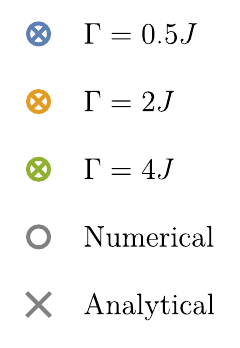} 
   }
   
   \caption{Comparison between the dissipation rates $\gamma_{i}$ calculated from \cref{eq:ri,eq:eigenvalues} versus that obtained by direct  numerical diagonalization of the dynamical matrix $A_{nm}$.  Results are shown for three systems:  \protect\subref{fig:DisSpec-1D-side} a one-dimensional chain with the drain attached to an edge site, \protect\subref{fig:DisSpec-1D-step} a one-dimensional chain with an antisymmetric step-function potential with strength $V = \pm 2J$ and the drain attached to the central site; \protect\subref{fig:DisSpec-Hoff} a two-dimensional square lattice with flux of $\varphi = \half[\pi]$ threaded through each plaquette and the drain attached to the top central site. 
   In all cases the lattice has $N=25$ sites and nearest-neighbor hopping (strength $J$). 
   The three lattices are sketched out in the top row. In the second row, we plot the ratio $\gamma_{i}/\Delta_{i}$ for three different values of the dissipative coupling $\Gamma$ (see legend). The energy spacing, $\Delta_{i} = \p{\gve_{i+1}-\gve_{i-1}}/2$, which is proportional to the group velocity, is plotted in \cref{fig:DissSpect} for these lattices.
   The circles are calculated by diagonalizing the dynamical matrix (see \cref{sec:model}). The crosses are calculated from \cref{eq:ri,eq:eigenvalues} using properties of the non-dissipative ($\Gamma = 0$) system only. We see remarkable agreement. We observe that while increasing the coupling from $\Gamma=0.5J$ to $\Gamma=2J$ increases the relaxation rate of all modes, in these cases a further increase to $\Gamma=4J$ creates Zeno-like suppression for most modes as the $\gve_{i}=0$ gains a macroscopic relaxation rate.
   }
   \label{fig:DissSpectNum}
\end{figure}

To test its validity, we have directly compared the dissipation spectrum calculated from the approximate expression in \cref{eq:eigenvalues} against an explicit numerical diaonalization of the dynamical matrix for a range of 1D and 2D models.  Representative results are shown in \cref{fig:DissSpectNum}.  One finds a very good agreement with the approximate dissipation rates. 
We find that $\gamma_{i}$ retains the qualitative behavior discussed above, including the perturbative regime at small $\Gamma$ and Zeno behavior at large $\Gamma$. \Cref{eq:eigenvalues}, however, gives us a full picture of the relaxation spectrum's behavior, as well as the transition between the two regimes.  Using this expression, we find:
\begin{equation}
\gamma_{i} \approx \fopt{\frac{\p{\Delta_{i}/\pi}^{2}}{\p{\Delta_{i}/\pi}^{2} + \p{r_{i}\Gamma/2}^{2}}\bar\Gamma_{i} & \bar \Gamma_{i}\ll \Delta_{i}
\\  \frac{\abs{\psi_{i}\br{\vn_{0}}}^{4}}{\abs{\psi_{i}\br{\vn_{0}}}^{4} + r_{i}^{2}}\tfrac{4}{\pi^{2}}\Delta_{i}^{2}/\bar\Gamma_{i}& \bar \Gamma_{i}\gg \Delta_{i}.
\\  \bar\Gamma_{i}\ln\br{1 + 4 \abs{\psi_{i}\br{\vn_{0}}}^{4}/r_{i}^{2}}^{1/4}& \half[\bar\Gamma_{i}] = \tfrac{\Delta_{i}}{\pi}.
}
\end{equation}
We see that for modes with $\abs{r_{i}}\ll \abs{\psi_{i}\br{\vn_{0}}}^{2}$, there is a logarithmic enhancement of in the relaxation rate at $\half[\bar\Gamma_{i}] = \tfrac{\Delta_{i}}{\pi}$. 

\begin{figure}[h] 
   \centering
   \parbox[c][0.22\textwidth][b]{0.05\textwidth}{
	\rotatebox{90}{$\gamma_{i}/\bar\Gamma_{i}$}
	}
   \subfloat[\label{fig:DisSpec-1D-side}1D chain, drain at end]
   {\parbox{0.25\textwidth}{
	\parbox[c][0.18\textwidth][c]{0.25\textwidth}{\includegraphics[width=0.25\textwidth]{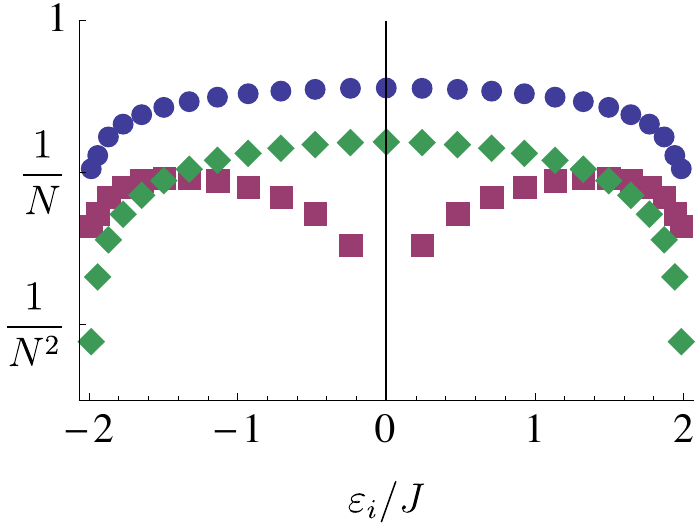}}
	
	\parbox[c][0.18\textwidth][c]{0.25\textwidth}{\includegraphics[width=0.25\textwidth]{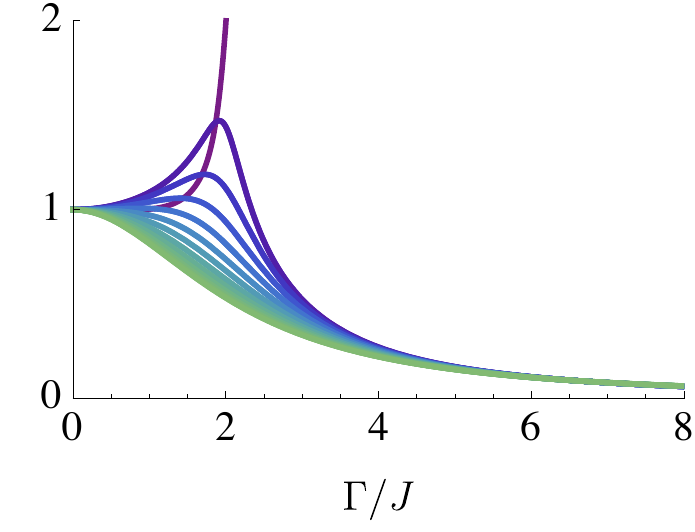}}
   }}
   \hfill
   \subfloat[\label{fig:DisSpec-1D-step}1D chain with step potential, drain at center]
   {\parbox{0.25\textwidth}{
	\parbox[c][0.18\textwidth][c]{0.25\textwidth}{\includegraphics[width=0.25\textwidth]{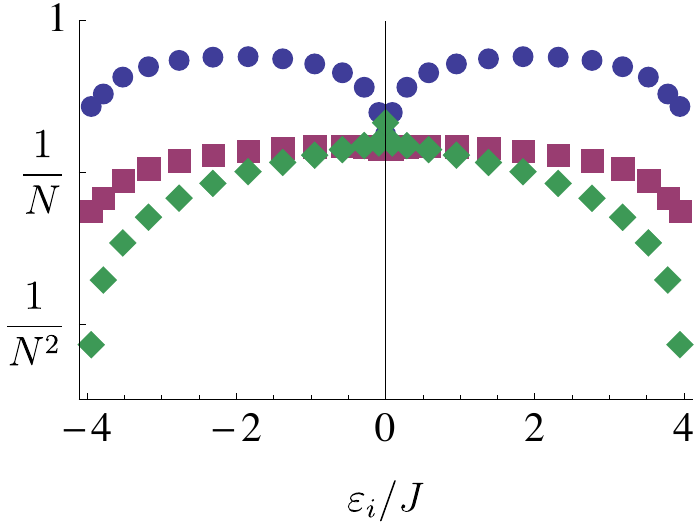}}
	
	\parbox[c][0.18\textwidth][c]{0.25\textwidth}{\includegraphics[width=0.25\textwidth]{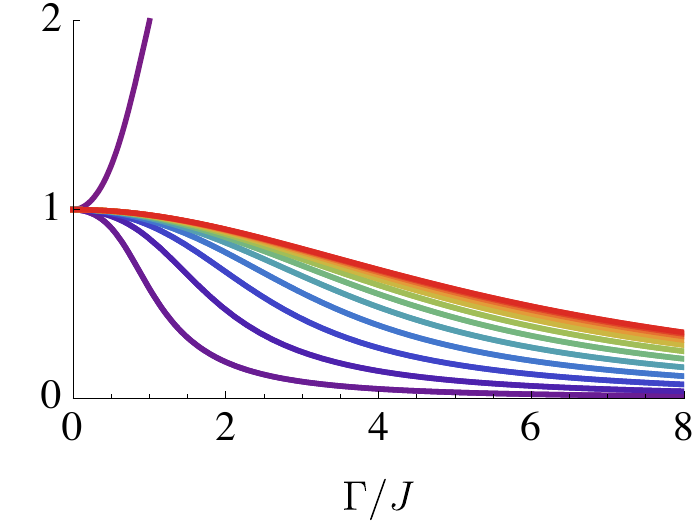}}
   }}
   \hfill
   \subfloat[\label{fig:DisSpec-Hoff}2D Hofstadter model with $\pi/2$ flux through each plaquette]
   {\parbox{0.25\textwidth}{
	\parbox[c][0.18\textwidth][c]{0.25\textwidth}{\includegraphics[width=0.25\textwidth]{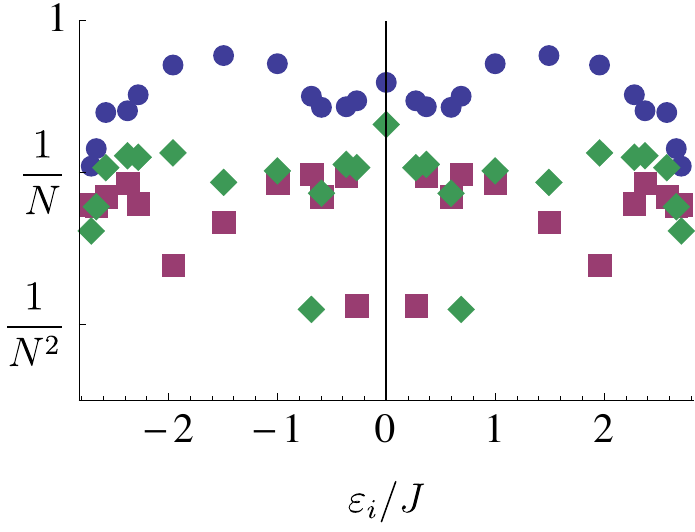}}
	
	\parbox[c][0.18\textwidth][c]{0.25\textwidth}{\includegraphics[width=0.25\textwidth]{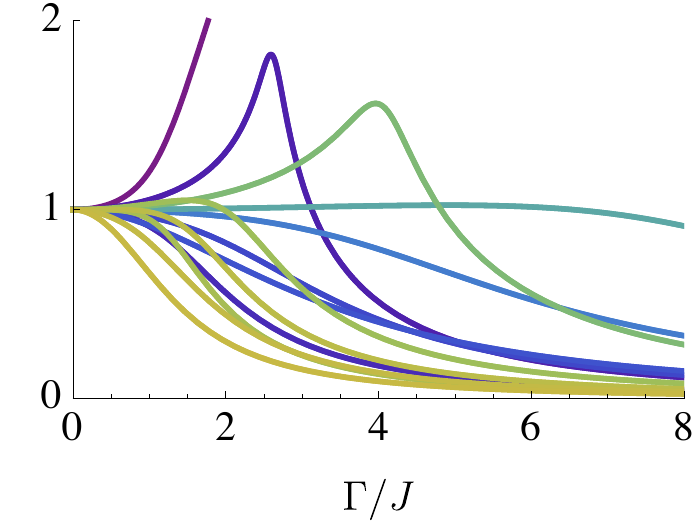}}
   }}
   \hfill
   \parbox[c][0.42\textwidth][c]{0.1\textwidth}{
	\parbox[c][0.2\textwidth][c]{0.1\textwidth}{\includegraphics[scale=1]{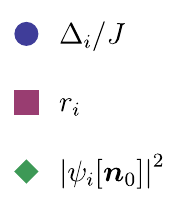}}
	
	\parbox[c][0.2\textwidth][c]{0.1\textwidth}{\includegraphics[scale=1]{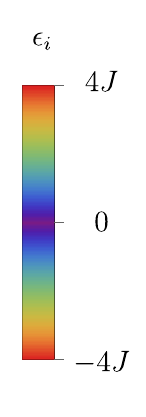}  }
   }
   
   \caption{Dependence of the dissipation spectrum on the coupling $\Gamma$. Plotted for the same systems shown in \cref{fig:DissSpectNum}.
   In the top row we plot the relevant parameters of the system's coherent energy eigenmodes appearing in \cref{eq:eigenvalues}, including, the local energy spacing $\Delta_{i}$, the calculated remainder $r_{i}$ (see \cref{eq:ri}), and coupling to the drain site, $\abs{\psi_{i}\br{\vn_{0}}}^{2}$. Note these are independent of the strength of the dissipation $\Gamma$. 
   In the bottom row we plot the calculated dissipation spectrum's variation as we vary $\Gamma$. The color of each line corresponds to the unperturbed energy $\gve_{i}$ of each mode (see color bar).
   We observe the behavior discussed near \cref{eq:gammaregimes}: for $\Gamma \lesssim J$, we are in the perturbative regime and the local relaxation rate is proportional to the coupling to the drain, $\gamma_{i}\sim \bar\Gamma_{i}$. As we increase $\Gamma$, the relaxation rate saturates near $\Gamma_{i}/2\approx \Delta_{i}/\pi$, and at higher values it is suppressed by quantum Zeno physics. The intermediate behavior is determined by the remainder $r_{i}$: where $\abs{r_{i}}< \abs{\psi_{i}\br{\vn_{0}}}^{2}$ (e.g.~in the center of the band in \protect\subref{fig:DisSpec-1D-side}) we see an initial enhancement of $\gamma_{i}$, while for larger $r_{i}$, elsewhere, the relaxation rate is monotonically suppressed.
   }
   \label{fig:DissSpect}
\end{figure}

It is also interesting to consider the ratio $\gamma_{i}/\bar \Gamma_{i}$ as a function of $\Gamma$; this ratio measures how different a mode's dissipation rate is from the simple, Fermi's Golden Rule estimate. For ${\abs{r_{i}} \geq \abs{\psi_{i}\br{\vn_{0}}}^{2}/\sqrt{3}}$, the ratio is a monotonically decreasing function of $\Gamma$. In this case, the mode essentially transitions directly from the perturbative regime to the Zeno regime. For $\abs{r_{i}} < \abs{\psi_{i}\br{\vn_{0}}}^{2}/\sqrt{3}$, however, $\gamma_{i}/\bar \Gamma_{i}$ initially increases, signifying the kind of impedance matching discussed above. Examples of this behavior are shown in \cref{fig:DissSpect} for several systems. We see the evolution of the spectrum take on quite different forms for different systems, with resonant enhancement appearing for none of the modes, for several modes at the same coupling strength, or for different modes at different values of the coupling.

\subsection{Dissipation spectrum of a flux ring \label{sec:fluxring}}

\begin{wrapfigure}[16]{L}{2in}
   \centering
   \includegraphics[width=2in]{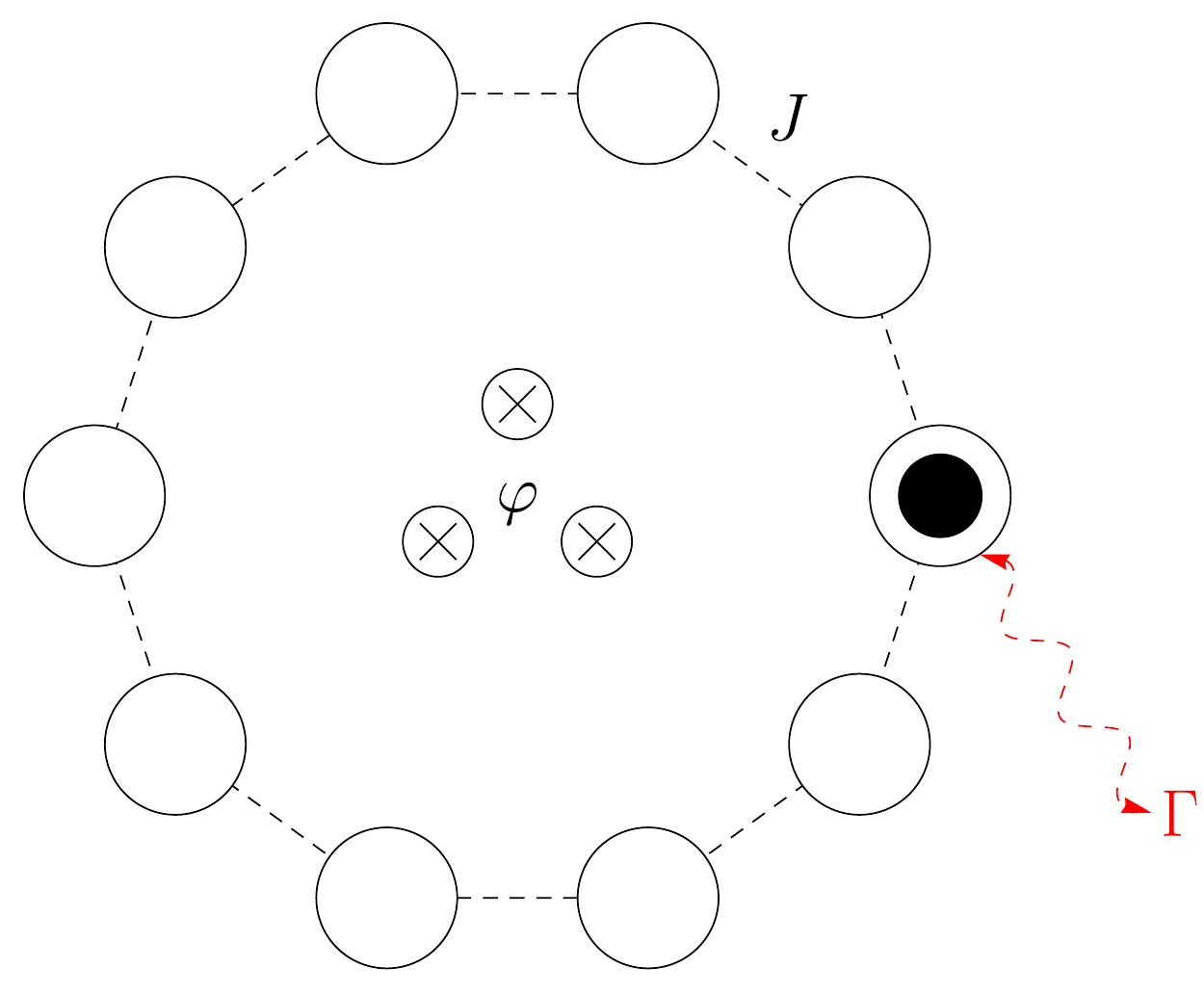} 
   \caption{A one-dimensional ring with $N=10$ sites and flux $\phi$ threaded through it, see \cref{eq:H1Dring}. A single drain site is coupled to a squeezed reservoir with dissipation rate $\Gamma$.}
   \label{fig:ring}
\end{wrapfigure}
We now apply the results of the previous section to a specific model of a one dimensional ring with nearest neighbour hopping (\cref{fig:ring}), pierced by a non-zero flux.  In this case, we can use \cref{eq:eigenvalues} to analytically calculate the dissipation spectrum to leading order in system size.  

The system Hamiltonian in this case is given by:
\begin{equation}\begin{gathered}
\hat H = -J\sum_{n=1}^{N}e^{-i\varphi/N}\hat a_{n+1}\dg\hat a_{n} + e^{i\varphi/N}\hat a_{n}\dg\hat a_{n+1}.
\label{eq:H1Dring}
\end{gathered}\end{equation}
We further assume periodic boundary conditions, $\hat a_{N+1}\equiv \hat a_{1}$, and restrict the ring flux $\varphi$ to the interval 
$[0,\pi ]$.
The system is diagonal in a plane-wave (i.e.~momentum) basis, 
\begin{equation}
\hat H = \sum_{k}\gve_{k}\hat b_{k}\dg\hat b_{k}, \qquad \gve_{k} = -2J\cos\p{k + \varphi/N},
\end{equation}
where $k = \tfrac{2\pi}{N}\times 0\dotsc N-1$. To order the modes so that $\gve_{i}\le \gve_{i+1}$, we relabel
\begin{equation}
\hat b_{k} \to \hat b_{i\br{k}} 
\qquad i\br{k} = -\half[N+1] + \fopt{ k/\tfrac{\pi}{N} & k \le \pi, \\ \p{2\pi - k}/\tfrac{\pi}{N} - 1 & k > \pi.}
\end{equation}
Labelled this way, every two consecutive eigenmodes have opposite wave-numbers, and in the absence of a flux (i.e.~$\varphi = 0$), are degenerate. These modes have energies
\begin{equation}
\hat H = \sum_{\mathclap{i=-\p{N-1}/2} }^{\mathclap{\p{N-1}/2}}\gve_{i}\hat b_{i}\dg\hat b_{i}, 
\qquad \gve_{j} = 2J\sin\br{\frac{\pi}{N}i + \p{-1}^{\p{i+\half[N-1]}}\frac{1}{N}\p{\varphi - \half[\pi]}}.
\end{equation}

To simplify the calculation we choose $\varphi = \half[\pi]$, where
\begin{equation}
\gve_{i} = 2J\sin \tfrac{\pi i}{N}, \qquad \Delta_{i} = \sin\tfrac{\pi}{N} 2J \cos \tfrac{\pi i}{N}.
\label{eq:1Dringspectrum}
\end{equation}
We note that due to translational invariance, $\abs{\br{\hat b_{j},\hat a_{n}\dg}}^{2} = 1/N$ for any $j,n$ and so for any choice $\vn_0$ for the position of the drain site, 
\begin{equation}
	\bar \Gamma_{i} = \frac{\Gamma}{N}.
\end{equation}
Translational invariance guarantees that all modes couple equally to the drain site.  

We can now calculate the remainder of \cref{eq:ri}, taking $N\gg 1$,
\begin{equation}
r_{j}  = \frac{1}{\pi}\sum_{l\ne j}\frac{\sin\tfrac{\pi}{N} 2J \cos \tfrac{\pi j}{N}}{2J\sin \tfrac{\pi j}{N} - 2J\sin \tfrac{\pi l}{N}}\frac{1}{N}
	 \approx \frac{\cos \tfrac{\pi j}{N}}{\pi N}\intrml{dx}{-\pi/2}{\pi/2}\frac{1 }{\sin \tfrac{\pi j}{N} - \sin x} = -i\frac{1}{N},
\end{equation}
We find, in this case, $r_{i} = O\p{\tfrac{1}{N}}$. Note that this is a feature of the ring system, and not generically true, as we have seen for different systems in \cref{fig:DissSpect}.

\begin{figure}[t] 
   \centering
   \parbox[b][0.12\textwidth][t]{0.05\textwidth}{
	\rotatebox{90}{$\gamma_{i}\quad\br{J/N}$}
	}
   \hfill
   \subfloat[\label{fig:1DgamJ1}$\Gamma=J$]{\includegraphics[width=0.25\columnwidth]{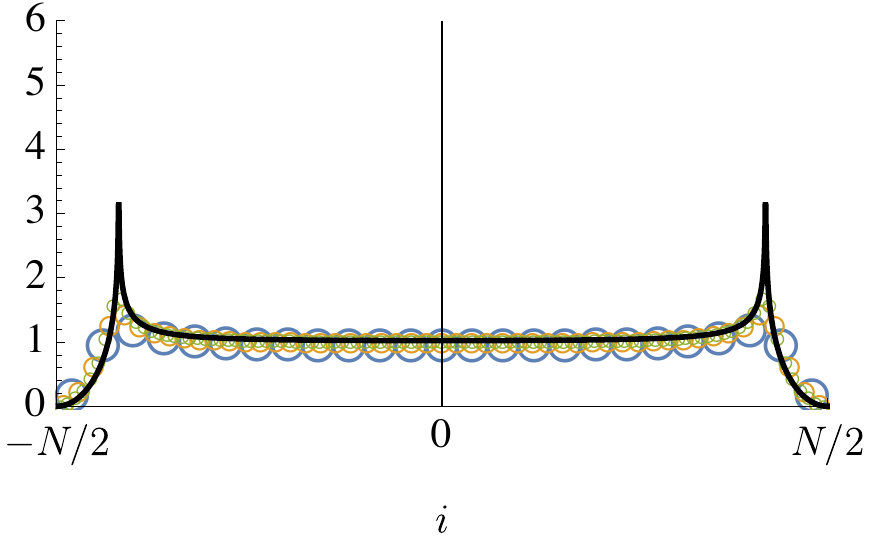} }
   \hfill
   \subfloat[\label{fig:1DgamJ3}$\Gamma=3J$]{\includegraphics[width=0.25\columnwidth]{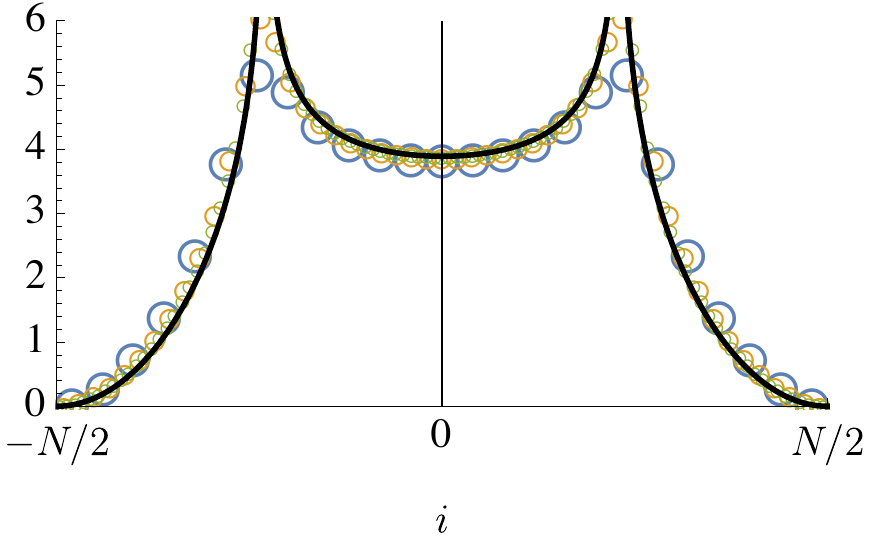} }
   \hfill
   \subfloat[\label{fig:1DgamJ5}$\Gamma=5J$]{\includegraphics[width=0.25\columnwidth]{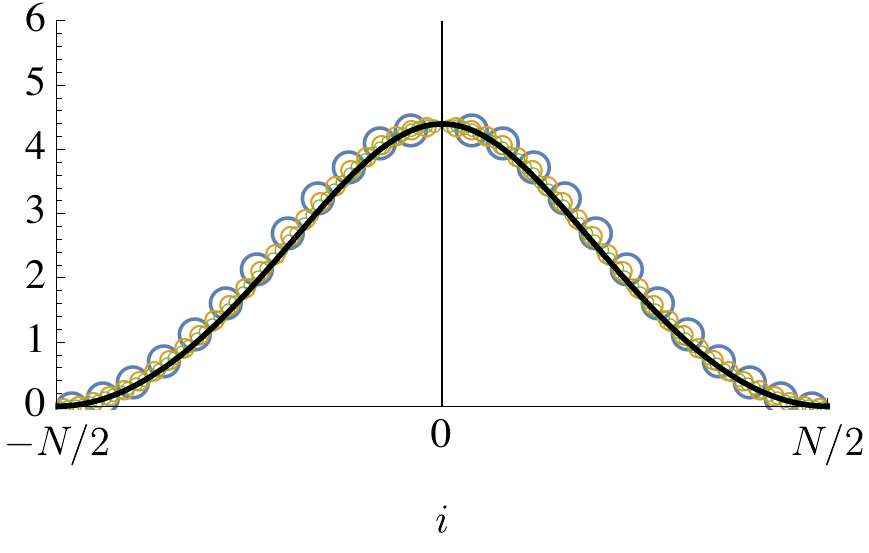} }
   \hfill
   \parbox[b]{0.15\textwidth}{
   \includegraphics[]{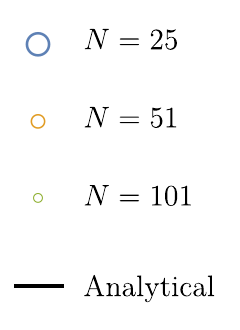} }
   \caption{The frequency shift  and dissipation spectrum of a one-dimensional ring with flux $\varphi = \half[\pi]$ threaded through (see \cref{eq:H1Dring} and \cref{fig:ring}) coupled to an external bath at a single site at different rates $\Gamma$. Here, the circles show the result from a numerical calculation for a finite system with different number of sites $N$, while the line is the analytical result for the $N\gg 1$ limit, given in \crefrange{eq:kc}{eq:overdampeddnu}.
    }
   \label{fig:1Dring}
\end{figure}

We next calculate the dissipation spectrum to first order in $1/N$. For $\Gamma < 4J$, we find a critical momentum, where the group velocity matches the dissipation rate,
\begin{equation}
2J\cos k_{c} = \half[\Gamma].
\label{eq:kc}
\end{equation}
For the relaxation rate, we see an impedance-matching like phenomenon, as discussed above, at this point,
\begin{equation}
\gamma_{i} \approx \fopt{ \frac{2J}{N}\cos \tfrac{\pi i}{N}\ln\abs{\frac{\cos \tfrac{\pi i}{N} + \cos k_{c}}{\cos \tfrac{\pi i}{N} - \cos k_{c}}} & \abs{\tfrac{\pi i}{N}} \ne k_{c},
	\\  \frac{2J}{N}\cos \tfrac{\pi i}{N} \ln N & \abs{\tfrac{\pi i}{N}} = k_{c}. }
\label{eq:underdampedgamma}
\end{equation}
In the overdamped case, $\Gamma > 4J$, there is no impedance matching. Instead, we see the relaxation rates dropping off away from the middle of the spectrum,
\begin{equation}
 \gamma_{i} = \frac{2J}{N}\cos \tfrac{\pi i}{N}\ln\frac{\Gamma + 4J\cos \tfrac{\pi i}{N}}{\Gamma - 4J\cos \tfrac{\pi i}{N}}.
 \label{eq:overdampedgamma}
\end{equation}
In this regime, we also find that \cref{eq:consist} is satisfied by a central rate with a macroscopic relaxation rate,
\begin{equation}
\gamma_{0} = \sqrt{\Gamma^{2} - 4J^{2}}.
 \label{eq:overdampeddnu}
\end{equation}
At large $\Gamma$, this mode becomes localized to the drain site, and effectively detaches from the rest of the ring.

These results are plotted in \cref{fig:1Dring}, along with a numerical calculation for finite systems of several sizes.

\section{Intermediate Time Behavior \label{sec:intertime}}

Having developed a full understanding of the dissipation spectrum associated with our local reservoir engineering setup, we now turn examining the more global features of the system's evolution from an initially prepared state to the final, dissipation-induced steady state.    

As described in \cref{sec:modes}, the eigenvalues $\gl_{i}$ of the dynamical matrix that characterize the system's time evolution can be usefully expressed as a dissipation-free energy plus a complex shift, $\gl_{i} = \gve_{i} + \gd\gl_{i}$; the imaginary part of this shift encodes the relaxation rate associated with a particular mode.   Our analysis revealed that these shifts, including the relaxation rate, are generally small and inversely proportional to system size $\gl_{i} \sim 1/N$. These basic features imply that the system's relaxation to the steady state can be broken into two parts.

\begin{itemize}

\item
First, for a relatively long period initial period (whose duration $\tau \propto N$), we can to a good approximation ignore the dissipation-induced contributions to the dynamical matrix eigenvalues, and simply replace them by the corresponding energy eigenvalue:
\begin{equation}
	e^{-i\gl_{i}t} \approx e^{-i\gve_{i}t} \quad \text{ for }\quad  t \ll 1/\gamma_{i} \sim N/J,
\label{eq:intert}
\end{equation}
where $J \propto \sum_{i}\Delta_{i}$ is the energy scale for the system's dynamics, i.e.~the hopping rate in the systems we consider.
During this initial period, the system evolution is well described by the non-dissipative dynamics generated by the system's coherent Hamiltonian.  At a heuristic level, particles can be injected into the system at the drain site, but will then propagate ballistically (i.e.~according to $\hat{H}$).  

\item
For longer times, the dissipative contribution to mode eigenvalues is non-neglible, and we have exponential decay associated with relaxation to the steady state:
\begin{equation}
	e^{-i\gl_{i}t} \approx e^{-i\gve_{i}t}e^{-\half[\gamma_{i}]t} \quad \text{ for } t \gtrsim 1/\gamma_{i} \sim N/J.
\label{eq:longt}
\end{equation}
At a heuristic level, this corresponds to a timescale long enough that particles injected from the drain site have had enough time to traverse the system and reflect off its boundaries.  This process continues for a long time until the steady state is reached.

\end{itemize}

We now attempt to see more explicitly how the above picture manifests itself in the state of the system.  
As the form of the dissipation-induced lattice state will depend on the form of the bath noise, we specify now to a Gaussian bath with a squeezed form. We take the input operator of \cref{eq:evolreal} to have correlators
\begin{equation}\begin{gathered}
\avg{\hat \zeta\dg\p{t}\hat \zeta\p{t\pr}} = \gd\p{t-t\pr} \Ns, \quad \avg{\hat \zeta\p{t}\hat \zeta\p{t\pr}} = \gd\p{t-t\pr} \Ms.
\end{gathered}\end{equation}

To understand the intermediate dynamics, we return to \cref{eq:evolbe} for the evolution of an energy eigenmode, rewriting it as
\begin{equation}
\dot{\hat b}_{i}\p{t} = -i\gve_{i}\hat b_{i}\p{t} + \psi_{i}\br{\vn_{0}}^{*}\p{\sqrt{\Gamma}\hat \zeta\p{t} - \half[\Gamma]\hat a_{\vn_{0}}\p{t}}
\label{eq:evolbwa0}
\end{equation}

During the initial evolution period, the simplest approximation would be to neglect the damping effect of the bath (last term) and only keep the driving term.  This would then correspond to a picture where the bath simply drives the system with correlated pairs of particles, but does not modify its dynamics or response properties.  

To get a slightly more accurate approximation of this early-period evolution, we can instead exactly solve the above equation for  
$\hat b_{i}\p{t}$ in terms of $\hat a_{\vn_{0}}$:
\begin{equation}
\hat a_{\vn_{0}}\p{t} = \sum_{i}\psi_{i}\br{\vn_{0}}\hat b_{i}\p{t} =  \sum_{i}\psi_{i}\br{\vn_{0}}e^{-i\gve_{i}t}\hat b_{i}\p{0} 
+ \intrml{d\tau}{0}{t}\br{\sum_{i}\abs{\psi_{i}\br{\vn_{0}}}^{2}e^{-i\gve_{i}\p{t-\tau}}}\p{\sqrt{\Gamma}\hat \zeta\p{\tau} - \half[\Gamma]\hat a_{\vn_{0}}\p{\tau}}.
\label{eq:atofa}
\end{equation}
The last two terms describe the driving and damping of the drain site by the bath as it is modified by the response of the rest of the lattice (e.g.~fluctuations may enter, bounce around the lattice several times, and then finally emerge at $\vec n_{0}$). This response is non-Markovian at short time scales, when these dynamics are sensitive to the finite bandwidth of the lattice; and at long times, when the discrete, non-uniform density of states  in the lattice is significant. However, during the intermediate regime of \cref{eq:intert}, a Markovian approximation is sufficient to qualitatively capture this effect.  We thus take 
\begin{equation}
	\sum_{i}\abs{\psi_{i}\br{\vn_{0}}}^{2}e^{-i\gve_{i}\p{t-\tau}} \approx \frac{1}{\J}\gd\p{t}
	\quad \text{ for } 1/J\ll  t \ll 1/\gamma_{i} \sim N/J
\label{eq:aapprox}
\end{equation}
where $\J\propto J$ is some factor dependent on the effective density of states at the drain.

Combining \cref{eq:evolbwa0,eq:atofa,eq:aapprox}, we arrive at an effective equation of motion,
\begin{equation}
\dot{\hat b}_{i}\p{t} = -i\gve_{i}\hat b_{i}\p{t} +\frac{\psi_{i}\br{\vn_{0}}^{*} }{1 + \Gamma/4\J}
	\p{\sqrt{\Gamma}\hat \zeta\p{t} - \frac{\Gamma}{2}\sum_{i}\psi_{i}\br{\vn_{0}}e^{-i\gve_{i}t}\hat b_{i}\p{0} }.
\end{equation}
This equation shows simple linear dynamics with a source term. Each mode is coupled to the source via its effective dissipation, with an overall suppression as the drain site $\hat a_{\vn_{0}}$ detaches from the rest of the lattice at large $\Gamma$. This additional factor captures some of the effect of the change in the system's dynamical eigenmodes, given in \crefrange{eq:eigenA}{eq:evoleigen}. These are significantly different from the system's original eigenmodes even at short and intermediate times.

The effective equation can be immediately solved. Taking the initial state to be the vacuum for simplicity, we find
\begin{subequations}\label{eq:corr2oft}\begin{gather}
\avg{\hat b_{i}\dg\p{t}\hat b_{j}\p{t}} = \frac{\Gamma \Ns}{\p{1 +  \Gamma/4\J}^{2}} 
	\psi_{j}\br{\vn_{0}}^{*} \psi_{i}\br{\vn_{0}}\frac{1 - e^{-i\p{\gve_{j}-\gve_{i}}t}}{i\p{\gve_{j} - \gve_{i}}},
\\ \avg{\hat b_{i}\p{t}\hat b_{j}\p{t}} = \frac{\Gamma\Ms}{\p{1 +  \Gamma/4\J}^{2}}
	\psi_{j}\br{\vn_{0}}^{*} \psi_{i}\br{\vn_{0}}^{*}\frac{1 - e^{-i\p{\gve_{j}+\gve_{i}}t}}{i\p{\gve_{j} + \gve_{i}}}.
\end{gather}\end{subequations}
We observe that the two-mode correlations depend on their energy difference, and evolves in multiple stages:
\begin{itemize}
\item Initially, while $\abs{\gve_{i}-\gve_{j}}t \lesssim 1$, the correlation grows linearly, $\avg{\hat b_{i}\dg\p{t}\hat b_{j}\p{t}} \propto t$.
\item After a time approximately equal to the corresponding rate, $\abs{\gve_{i}-\gve_{j}}t \gtrsim 1$, the correlation saturates with magnitude $\avg{\hat b_{i}\dg\p{t}\hat b_{j}\p{t}} \propto 1/\p{\gve_{i} - \gve_{j}}$.
\item For the duration of the intermediate time regime, until $t\sim N/\Gamma$, these correlations are then independent of time up to a rotating term.
\item Finally, at long times, further equilibration occurs, associated with the dissipative contribution to the dynamical mode eigenvalues (which are neglected here).   
\end{itemize}
The anomalous correlations behave similarly with regards to the sum of the energies.

\begin{figure}[t] 
   \centering

   \subfloat[\label{fig:1DMTbb2Dplot}
   ]{
	\parbox{0.54\textwidth}{
	   \includegraphics[width=0.17\columnwidth]{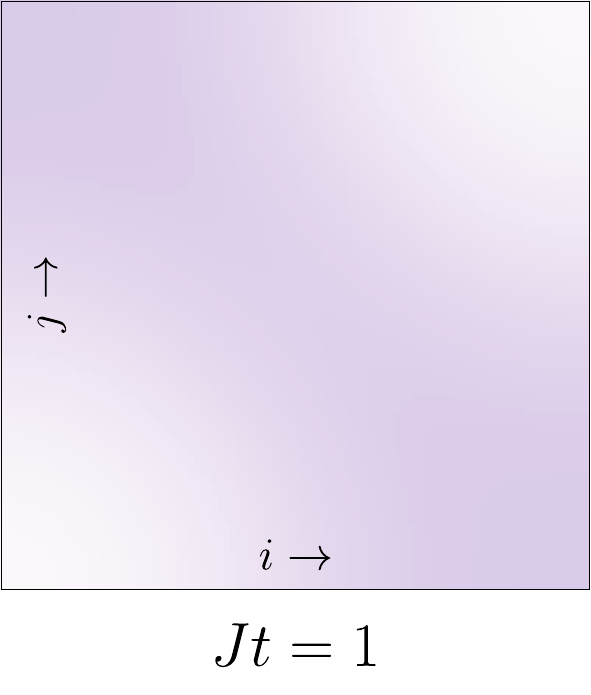} \hfill
	   \includegraphics[width=0.17\columnwidth]{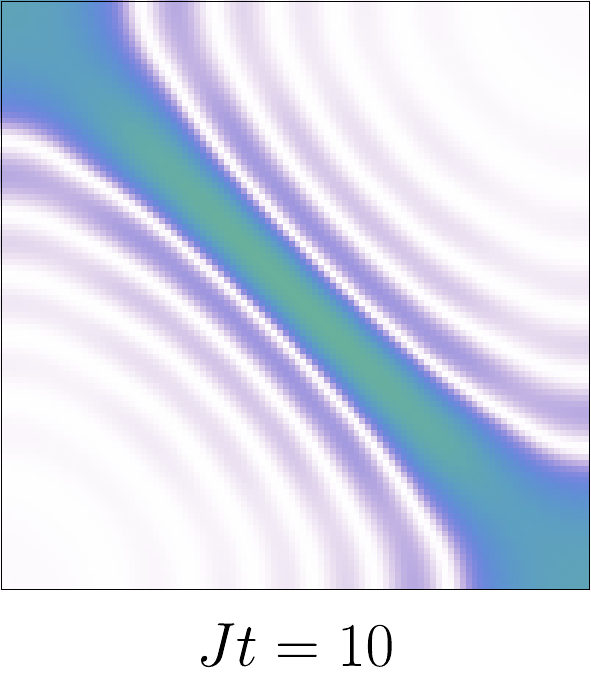} \hfill
	   \includegraphics[width=0.17\columnwidth]{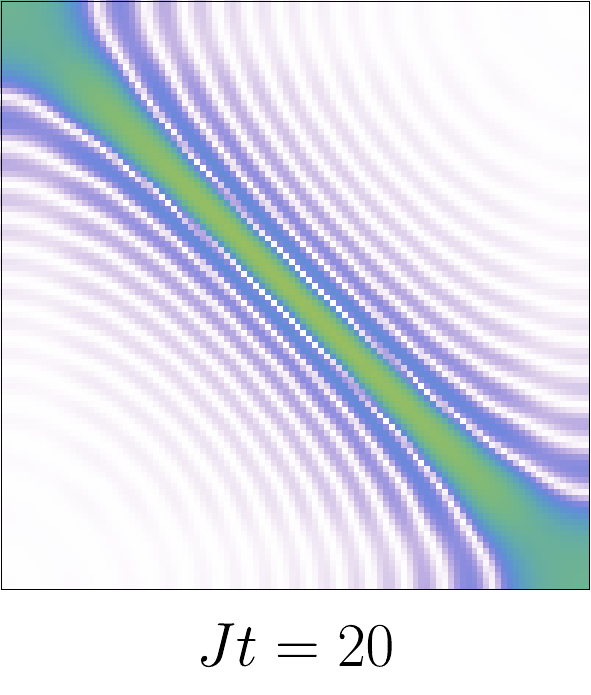} \hfill
	   
	   \vspace{10pt}
	   \includegraphics[width=0.17\columnwidth]{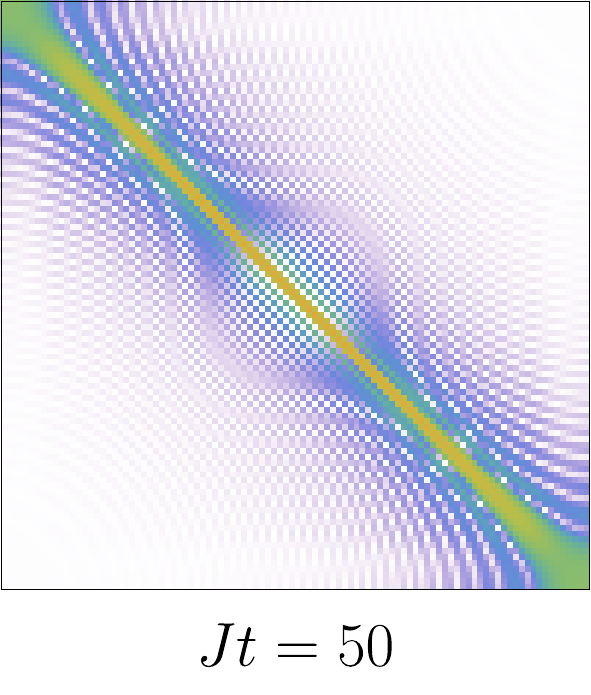} \hfill
	   \includegraphics[width=0.17\columnwidth]{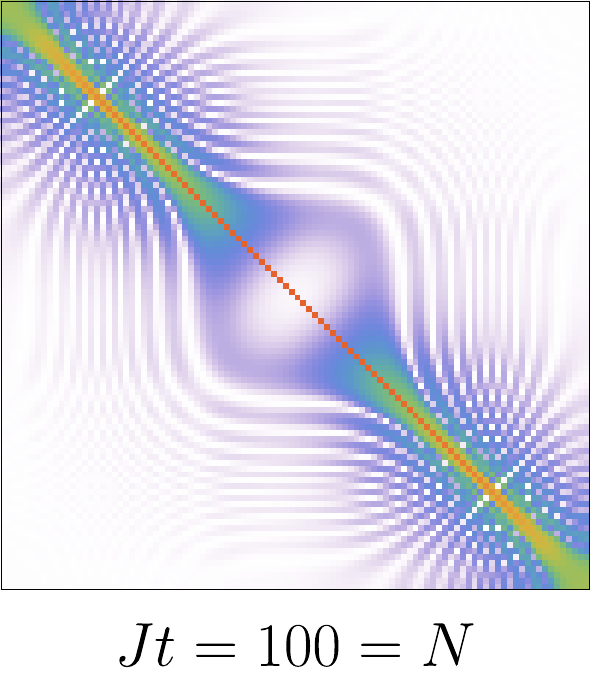} \hfill
	   \includegraphics[width=0.17\columnwidth]{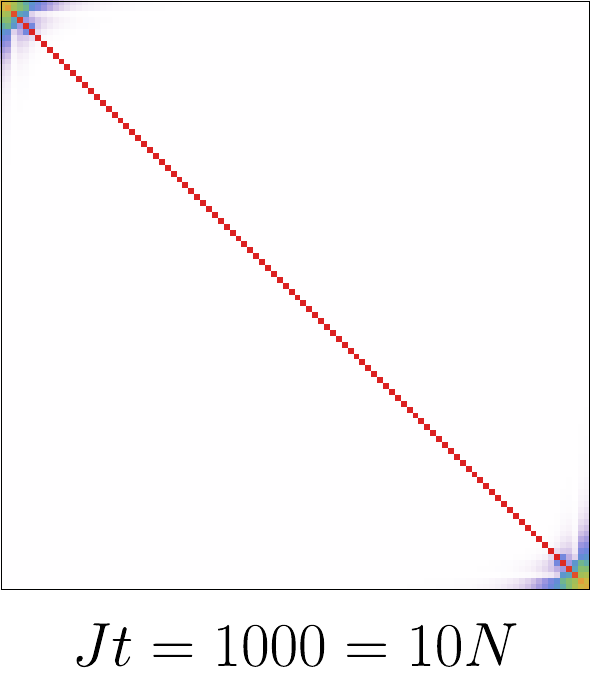} 
	  }
	\begin{tabular}{c}
		$\abs{\avg{\hat b_{i}\hat b_{j}}}^{2}/\mathcal M$
		\\ \includegraphics[height=0.34\columnwidth]{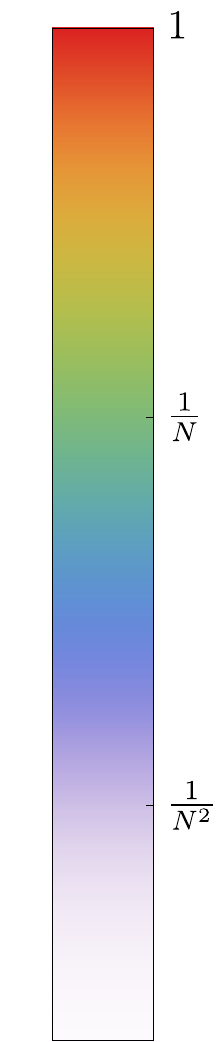}
		\\ $N=100$
	\end{tabular}
   }
   \parbox{0.3\textwidth}{
   \subfloat[\label{fig:1DMTbboftplot}]{
	   \includegraphics[width=0.3\textwidth]{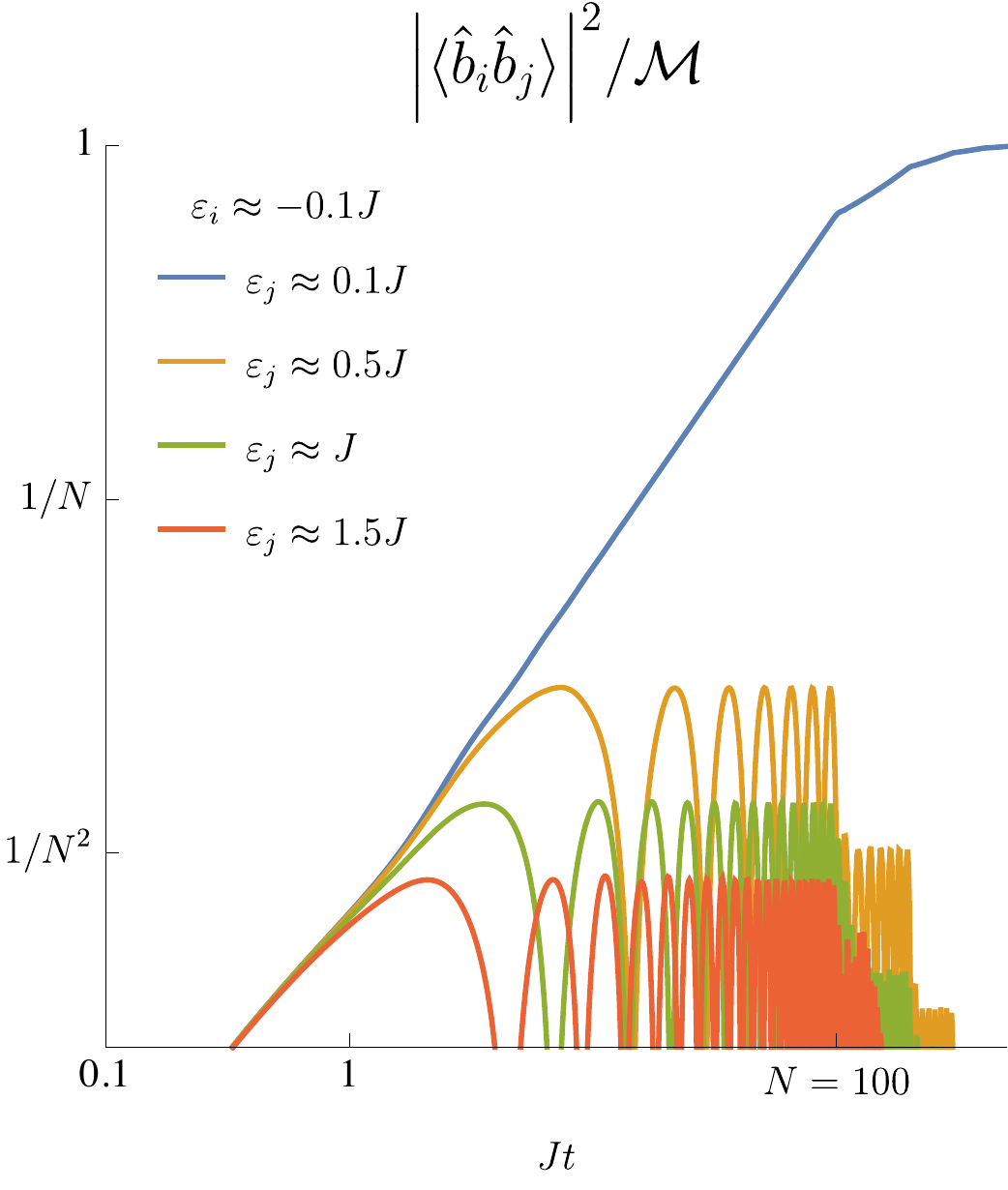} 
   }}

   \caption{Intermediate-time behavior of the energy eigenmode correlations in a one-dimensional ring with flux $\pi/2$ threaded through it (see \cref{eq:H1Dring,fig:ring}). Here $N=100$, $\Gamma=J$.
   	\protect\subref{fig:1DMTbb2Dplot} shows all correlations at specific times, while \protect\subref{fig:1DMTbboftplot} shows a select number of correlations at all times.
	We observe the behavior described by \cref{eq:corr2oft}: at short time, a linear growth of correlations, saturating with at $\avg{\hat b_{i}\hat b_{j}}\sim1/\abs{\gve_{i}-\gve_{j}}$. Then, during the intermediate regime, $1\ll Jt \ll N$, the correlations show oscillating behavior. At long times, $Jt \gtrsim N$, we observer the dissipative relaxation into the chiral steady state, $\avg{\hat b_{i}\hat b_{j}}\sim \gd_{i,-j}$.
   }
   \label{fig:1DringMedTime}
\end{figure}

In \cref{fig:1DringMedTime} we plot the intermediate-time behavior of the energy eigenmode correlations of the simple system described in \cref{sec:fluxring}, the one dimensional ring with flux threaded through it. We observe the behavior outlined above.

We can also find the intermediate time correlations in real space. Within the approximation of \cref{eq:aapprox}, they are
\begin{equation}\begin{split}
\avg{\hat a_{\vm}\dg\p{t}\hat a_{\vn}\p{t}} & = \sum_{i,j}\psi_{i}^{*}\br{\vm}\psi_{j}\br{\vn}\avg{\hat b_{i}\dg\p{t}\hat b_{j}\p{t}}
\\ &=  \frac{\Gamma \mathcal N}{\p{1 +  \Gamma/4\J}^{2}} \sum_{i,j}\psi_{i}^{*}\br{\vm}\psi_{i}\br{\vn_{0}}\times \psi_{j}^{*}\br{\vn_{0}}\psi_{j}\br{\vn}\times 
	\p{\frac{1 - e^{-i\p{\gve_{j}-\gve_{i}}t}}{i\p{\gve_{j} - \gve_{i}}}}.
\end{split}\end{equation}
If the modes of the system are extended and have a plane-wave character, a ballistic behavior pattern emerges. If we take the wavefunctions to behave as 
\begin{equation}
\psi_{i}^{*}\br{\vm}\psi_{i}\br{\vn_{0}} \propto e^{- i\gve_{i}\tau_{\vec m}}, \qquad \tau_{\vec m} = \abs{\vm-\vn_{0}}/c
\end{equation}
for some propagation speed $c$, then $\avg{\hat a_{\vm}\dg\p{t}\hat a_{\vn}\p{t}}$ is simply the Fourier transform of the term in parentheses, evaluated at $\tau_{\vec m},\tau_{\vec n}$. However, as these correlators are cut off at width $t$, while narrower features are largely defined by the time-independent post-saturation correlations. Thus, in the presence of a characteristic propagation speed $c$, we can expect a behavior of the form
\begin{equation}
\avg{\hat a_{\vm}\dg\p{t}\hat a_{\vn}\p{t}} = \fopt{ 0 & \max{\abs{\vm-\vn_{0}},\abs{\vn-\vn_{0}}} \lesssim ct
	\\ \text{Constant} & \abs{\vm-\vn_{0}},\abs{\vn-\vn_{0}} \gtrsim ct.}
\label{eq:amanintt}
\end{equation}

\begin{figure}[t] 
   \centering

   \subfloat[\label{fig:1DMTaa2Dplot}
   ]{
	\parbox{0.54\textwidth}{
	   \includegraphics[width=0.17\columnwidth]{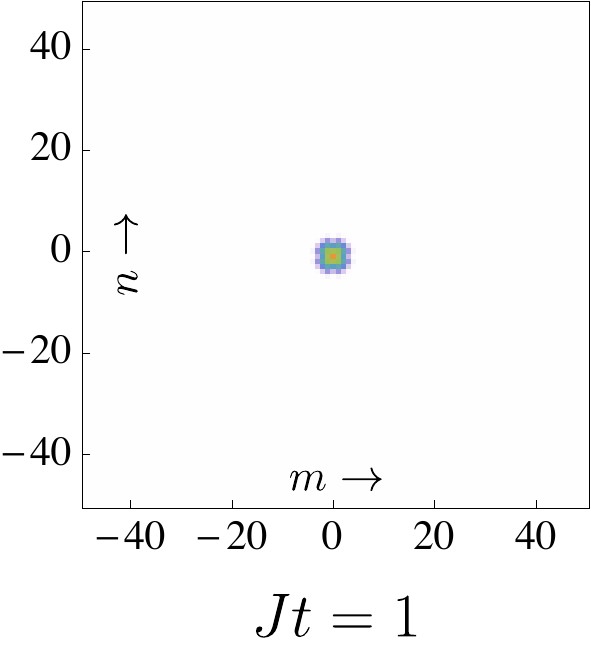} 
	   \includegraphics[width=0.17\columnwidth]{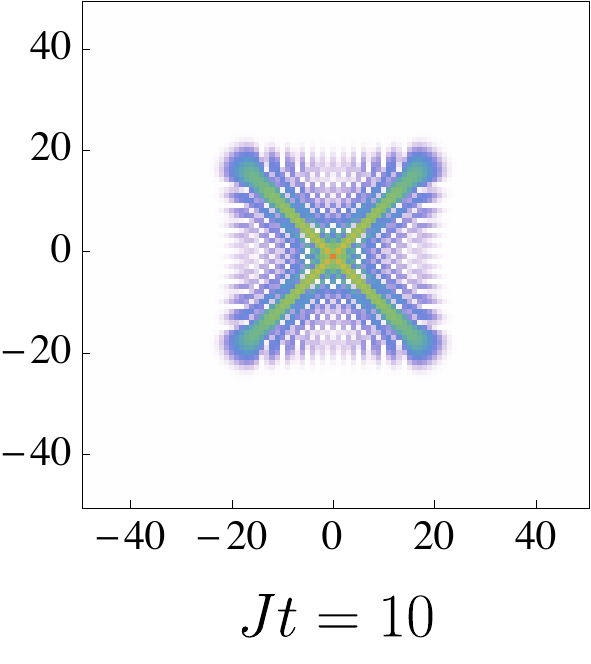} 
	   \includegraphics[width=0.17\columnwidth]{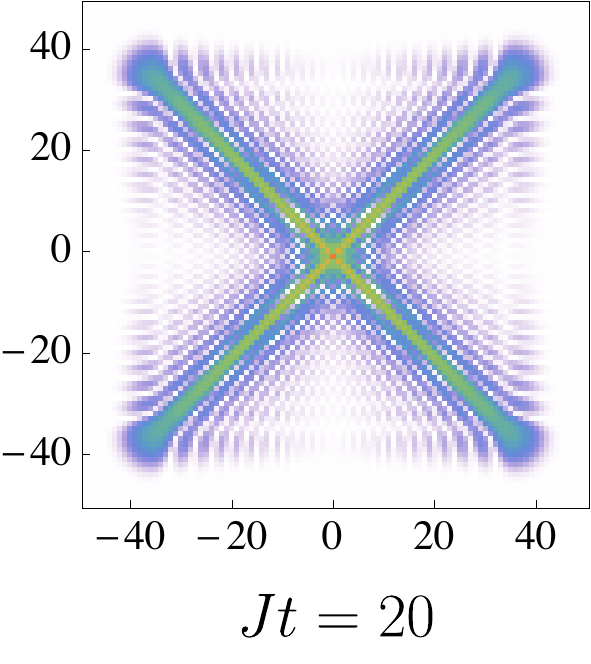} 
	   
	   \vspace{10pt}
	   
	   \includegraphics[width=0.17\columnwidth]{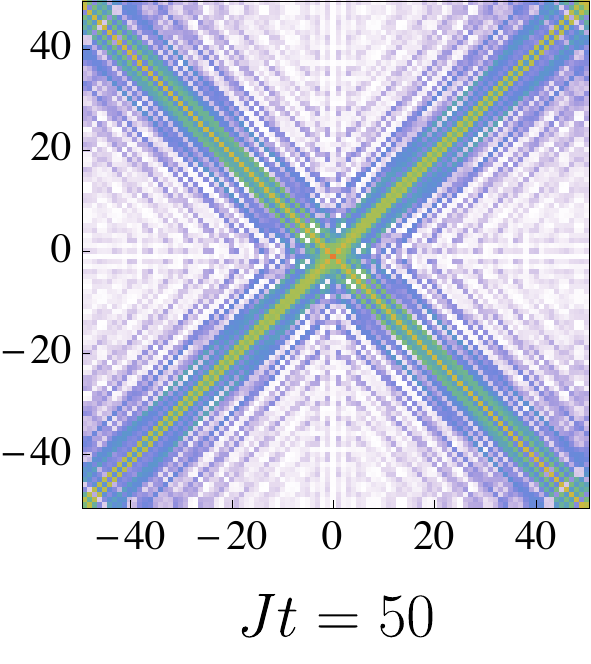} 
	   \includegraphics[width=0.17\columnwidth]{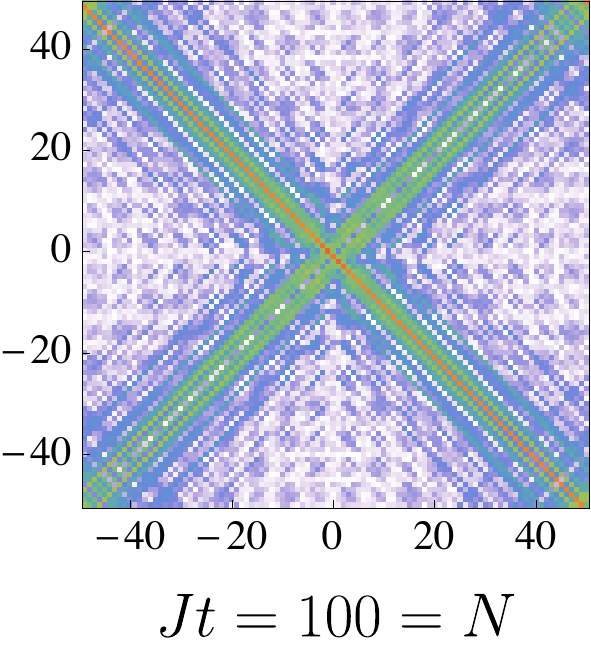} 
	   \includegraphics[width=0.17\columnwidth]{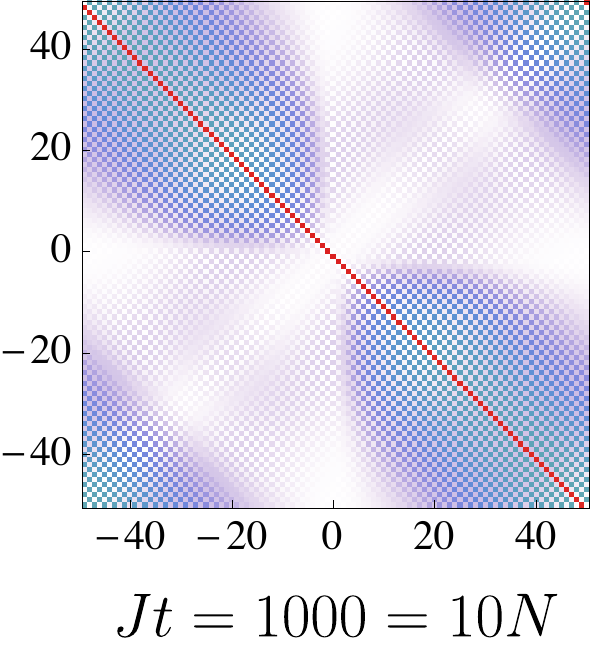} 
	  }
	\begin{tabular}{c}
		$\abs{\avg{\hat a_{m}\hat a_{n}}}^{2}/\mathcal M$
		\\ \includegraphics[height=0.34\columnwidth]{MedTimeleg.pdf}
		\\ $N =100$
	\end{tabular}
   }
   \parbox{0.3\textwidth}{
   \begin{tabular}{c}
   \subfloat[\label{fig:1DMTapapoftplot}Same-site correlations]{
	   \includegraphics[width=0.31\textwidth]{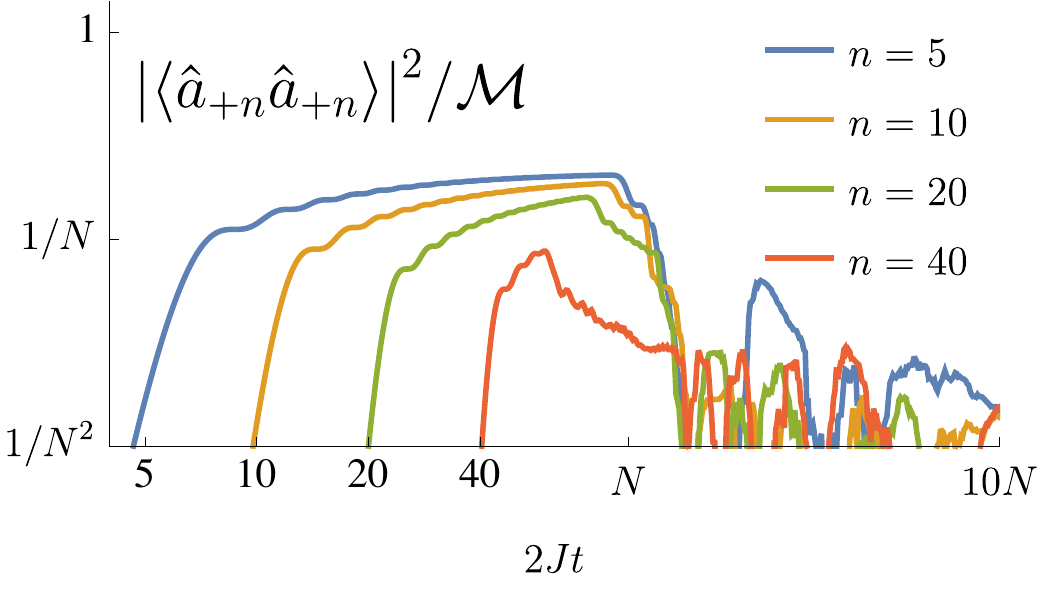}}
   \\
   \subfloat[\label{fig:1DMTapamoftplot}Mirror correlations]{
	   \includegraphics[width=0.31\textwidth]{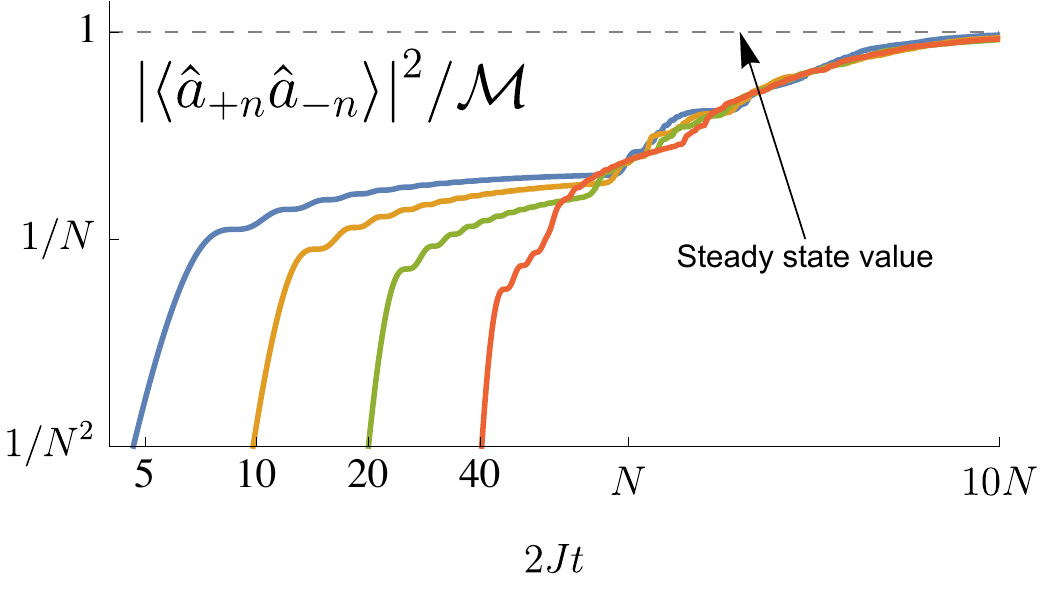}}
   \end{tabular}}

   \caption{Intermediate-time behavior of the energy eigenmode correlations. Here, shown for a one-dimensional ring with flux $\pi/2$ threaded through it and the drain site coupled at $n=0$ (see \cref{eq:H1Dring,fig:ring}). Here $N=100$, $\Gamma=J$.
   	\protect\subref{fig:1DMTaa2Dplot} shows all correlations at specific times, while \protect\subref{fig:1DMTapapoftplot}-\protect\subref{fig:1DMTapamoftplot} shows a select number of correlations at all times.
	We observe the ballistic behavior described in \cref{sec:intertime}: a distinctive light cone defined by the maximum velocity of the lattice ($v=2J$), with no correlations outside of it. It is interesting to compare the behavior of the same-site correlations, $\avg{\hat a_{n}\hat a_{n}}$, plotted in \protect\subref{fig:1DMTapapoftplot} to the mirror correlations, $\avg{\hat a_{n}\hat a_{-n}}$, plotted in \protect\subref{fig:1DMTapamoftplot}. In the intermediate regime, these behave similarly, coming into the light-cone at the same time and growing in the same way. In the long time regime, the dissipative dynamics drive this chiral system into a mirror-correlation steady state, and so the same-site correlations vanish while mirror correlations grow.
   }
   \label{fig:1DringMedTimeaa}
\end{figure}

We show this behavior for a one dimensional ring in \cref{fig:1DringMedTimeaa}, observing the behaviors outlined above. We observe a clear light-cone, with correlations cut off at a distance of $2Jt$ from the drain. Within this light cone, the correlations asymptotically approach a fixed pattern. Notably, this pattern is quite different from the steady state, exhibiting same-site anomalous correlations $\avg{\hat a_{n}\hat a_{n}} \sim \avg{\hat a_{n}\hat a_{-n}}$ which vanish in the long term.

\section{Conclusions}

We have analyzed here the dynamical properties of a bosonic lattice locally coupled to a single Markovian and Gaussian reservoir, with a particular focus on a squeezed bath. In calculating the the spectrum of its relaxation rates, we found that it is largely defined by the ratio $\bar\Gamma_{i}/\Delta_{i}$, crossing over from a perturbative coupling between the system's modes and the bath to a Zeno-like suppression when the drain is strongly coupled. We have also shown that the regime of intermediate-valued bath couplings can exhibit rich behavior, including resonant amplification of some modes' relaxation rates. Using \cref{eq:ri,eq:eigenvalues} this behavior can be analytically calculated for a given system, allowing these parameters to be chosen to e.g.~optimize the over relaxation time.

We have also explored the intermediate time behavior of such systems in the case of a large lattice, proving the existence of a distinct intermediate time regime.
There, we found that a pre-thermalized correlation pattern emerges, which may be quite different from both the initial state and steady state.

The local squeezing bath we suggest here could be experimentally realized as a modification of existing experiments \cite{Anderson2016,Owens2018}. As such, the dynamics we have calculated could be directly observed, and may prove useful, e.g.~in generating desired entanglement in microwave cavities. 
Further modifications, such coupling to a non-Markovian bath or the addition of particle interactions, may reveal even richer physics in such lattices.

\section*{Acknowledgements}

This work was supported by the Air Force Office of Scientific Research MURI program, under grant number FA9550-19-1-0399.

\appendix

\section{Time evolution eigenmodes\label{app:eigens}}

Our system evolves according to the master equation given in \cref{eq:evolbe}. 

We will show that the left-eigenmodes of $A$ are given by
\begin{equation}
V_{ij} = \frac{1}{2}\frac{e^{i\varphi_{j}}\sqrt{\bar\Gamma_{j}}}{\gl_{i} - \gve_{j}},
\end{equation}
where the eigenvalues $\gl_{i}$ are solutions of the self-consistency equation
\begin{equation}
S\p{\gl} = \frac{1}{2}\sum_{j}\frac{\bar\Gamma_{j}}{\gl - \gve_{j}} = i.
\end{equation}

These have
\begin{equation}\begin{split}
\nonumber
\sum_{j} V_{ij}A_{jl} & = \sum_{j}\frac{1}{2}\frac{e^{i\varphi_{j}}\sqrt{\bar\Gamma_{j}}}{\gl_{i} - \gve_{j}}
	\p{\gd_{j,l}\gve_{j} - i e^{i\p{\varphi_{l}-\varphi_{j}}}\half \sqrt{\bar \Gamma_{j}\bar \Gamma_{l}}}
	\\ & = \frac{1}{2}\frac{e^{i\varphi_{l}}\sqrt{\bar\Gamma_{l}}}{\gl_{i} - \gve_{l}}\gve_{l}
		- ie^{i\varphi_{l}}\frac{\sqrt{\bar \Gamma_{l}}}{2}\frac{1}{2}\sum_{j}\frac{\bar\Gamma_{j}}{\gl_{i} - \gve_{j}} 
	= \frac{1}{2}\frac{e^{i\varphi_{l}}\sqrt{\bar\Gamma_{l}}}{\gl_{i} - \gve_{l}}\gve_{l} + e^{i\varphi_{l}}\frac{\sqrt{\bar \Gamma_{l}}}{2}
	\\ & = \frac{1}{2}\frac{e^{i\varphi_{l}}\sqrt{\bar\Gamma_{l}}}{\gl_{i} - \gve_{l}}\gl_{i} = \gl_{i}V_{i,l}.
\end{split}\end{equation}
satisfying \cref{eq:lefteigenAdef}.

Then, from \cref{eq:evolbe} we find
\begin{equation}\begin{split}
\dot{\tilde b}_{i} = \sum_{j} V_{ij}\dot{\hat b}_{j} & = -i \sum_{j,l} V_{ij}A_{jl}\hat b_{l} + \sum_{j}V_{ij}e^{-i\varphi_{j}}\sqrt{\bar \Gamma_{j}}\hat\zeta
	\\ & =  -i \gl_{i}\sum_{l} V_{i,l}\hat b_{l} +\frac{1}{2}\sum_{j}\tfrac{\bar\Gamma_{j}}{\gl_{i} - \gve_{j}}\hat\zeta
	= -i\gl_{i}\tilde b_{i} + i\hat \zeta.
\end{split} \end{equation}
and so we may immediately integrate to find $\tilde b_{i}\p{t}$.

The inverse relation is given by
\begin{equation}\begin{gathered}
V^{-1}_{ij} = \frac{e^{-i\varphi_{i}} \sqrt{\bar \Gamma_{i}}G_{j}}{\gl_{j}-\gve_{i}},
	\qquad 1/G_{i} = \frac{1}{2}\sum_{j}\frac{\bar \Gamma_{j}}{\p{\gl_{i}-\gve_{j}}^{2}}.
\end{gathered}\end{equation}
This is shown by
\begin{equation}\begin{split}
\nonumber
\sum_{l}V_{il}V^{-1}_{lj} & = 
	\sum_{l}\frac{1}{2}\frac{e^{i\varphi_{l}}\sqrt{\bar\Gamma_{l}}}{\gl_{i} - \gve_{l}}\frac{e^{-i\varphi_{l}} \sqrt{\bar \Gamma_{l}}G_{j}}{\gl_{j}-\gve_{l}}
		= G_{j} \frac{1}{2}\sum_{l}\bar \Gamma_{l}\frac{1}{\gl_{i} - \gep_{l}}\frac{1}{\gl_{j} - \gep_{l}}
\\ & = G_{j}\br{\gd_{i,j}\frac{1}{2}\sum_{l}\frac{\bar \Gamma_{l}}{\p{\gl_{i} - \gve_{l}}^{2}}
	+ \frac{1 - \gd_{i,j}}{\gl_{i} - \gl_{j}}\frac{1}{2}\sum_{l}\p{\frac{\bar \Gamma_{l}}{\gl_{j} - \gep_{l}} - \frac{\bar \Gamma_{l}}{\gl_{i} - \gep_{l}}}}
\\ & = G_{j}\br{\gd_{i,j}/G_{i} + \frac{1 - \gd_{i,j}}{\gl_{i} - \gl_{j}}\p{i - i}}  = \gd_{i,j}.
\end{split}\end{equation}

Note that
\begin{equation}\begin{gathered}
\forall j\quad \dee{}{\bar\Gamma_{j}}S\p{\gl_{i}} = \frac{1}{2}\frac{1}{\gl_{i}-\gve_{j}} - \frac{1}{2}\sum_{l}\frac{\bar \Gamma_{l}}{\p{\gl_{i} - \gve_{l}}^{2}}\dee{\gl_{i}}{\bar\Gamma_{j}}
	= \frac{1}{2}\frac{1}{\gl_{i}-\gve_{j}} - \frac{1}{G_{i}}\dee{\gl_{i}}{\bar\Gamma_{j}}
\\   \dee{}{\bar\Gamma_{j}}S\p{\gl_{i}}  = 0
\qquad \Rightarrow G_{i} = 2\p{\gl_{i} - \gve_{j}}\dee{\gl_{i}}{\bar\Gamma_{j}},
\end{gathered}\end{equation}
and we can calculate
\begin{equation}\begin{gathered}
\sum_{j}V_{ij}^{-1} = e^{-i\varphi_{i}} \sqrt{\bar \Gamma_{i}}\sum_{j}\frac{G_{j}}{\gl_{j}-\gve_{i}}
	= e^{-i\varphi_{i}}\sqrt{\bar \Gamma_{i}} \sum_{j}2\dee{}{\bar\Gamma_{i}}\gl_{j}
\\ \sum_{j}\gl_{j} = \sum_{j}\p{\gve_{j} - i \half[\bar\Gamma_{j}]}
\\ \Rightarrow \sum_{j}V_{ij}^{-1} = -ie^{-i\varphi_{i}}\sqrt{\bar \Gamma_{i}}.
\end{gathered}\end{equation}

\clearpage

\end{document}